\documentclass[10pt,journal]{IEEEtran}
\usepackage{amsfonts}
\usepackage{algorithmicx,algorithm}
\usepackage{array}
\usepackage{textcomp}
\usepackage{url}
\usepackage{verbatim}
\usepackage{cite}

\usepackage{subfigure}
\usepackage[center]{caption2}
\usepackage{stfloats}
\usepackage{float}
\usepackage{times}
\usepackage{latexsym}
\usepackage{bm}
\usepackage{setspace}
\usepackage{color}
\usepackage{graphicx}
\usepackage{citesort}
\usepackage{multirow}
\usepackage{epstopdf}
\usepackage{amsthm,amsmath,amssymb,lipsum}
\usepackage{mathrsfs}
\allowdisplaybreaks[4]
\usepackage{algpseudocode}
\usepackage{fancyhdr}
\usepackage{subeqnarray}
\usepackage{cases}

\newtheorem{remark}{Remark}

\begin{document}
\title{ Location Sensing and Beamforming Design for IRS-Enabled Multi-User ISAC Systems}
\author{\IEEEauthorblockN{Zhouyuan Yu, Xiaoling Hu, {\em Member, IEEE},   Chenxi Liu, {\em Member, IEEE}, \\ Mugen Peng, {\em Fellow, IEEE}  and Caijun Zhong, {\em Senior Member, IEEE}}
\thanks{Zhouyuan Yu, Xiaoling Hu, Chenxi Liu  and Mugen Peng  are with the State Key Laboratory of Networking and Switching Technology, Beijing University of Posts and
Telecommunications, Beijing 100876, China (e-mail: \{yzy9912, xiaolinghu, chenxi.liu, pmg\}@bupt.edu.cn).}
\thanks{Caijun Zhong is with the College of Information Science and Electronic Engineering, Zhejiang University, Hangzhou 310027, China (email: caijunzhong@zju.edu.cn).}
}
\maketitle

\begin{abstract}
This paper explores the potential of the intelligent reflecting surface (IRS) in realizing multi-user concurrent communication and localization, using the same time-frequency resources. Specifically, we propose an IRS-enabled multi-user integrated sensing and communication (ISAC) framework, where a distributed semi-passive IRS assists the uplink data transmission from  multiple users to the base station (BS) and conducts multi-user localization, simultaneously. We first design an ISAC  transmission  protocol, where the whole transmission period  consists of two periods, i.e., the ISAC period for simultaneous uplink communication and multi-user  localization, and the pure communication (PC) period for only uplink data transmission. For the ISAC period, we propose a multi-user location sensing algorithm, which utilizes the uplink communication signals unknown to the IRS, thus removing the requirement of dedicated positioning reference signals in conventional location sensing methods. Based on the sensed users' locations, we propose two novel beamforming algorithms for the ISAC period and PC period, respectively, which can work with discrete phase shifts and require no channel state information (CSI) acquisition. Numerical results show that the proposed multi-user location sensing algorithm can achieve up to millimeter-level positioning accuracy, indicating the advantage of the IRS-enabled ISAC framework. Moreover, the proposed beamforming algorithms with sensed location information and  discrete phase shifts can achieve comparable performance to 
the benchmark considering perfect CSI acquisition and continuous phase shifts, demonstrating how the location information can ensure the communication performance.
\end{abstract}


\section{Introduction}
Intelligent reflecting surface (IRS) is considered as a promising technology for the sixth-generation (6G) mobile communications, owing to its potential of enhancing the capacity and coverage of wireless network by proactively reconfiguring the wireless propagation environment \cite{9424177}. In general, the IRS is an electromagnetic two-dimensional metasurface composed of a large array of passive reflecting elements, each of which can independently impose the required phase shift on the incident signal to create a desirable multi-path effect \cite{wu2019intelligent,gong2020toward,wu2019towards}. Moreover, the IRS can passively reflect the incident signals without any sophisticated signal processing operations that require radio-frequency (RF) transceiver hardware, thereby significantly reducing hardware cost and energy consumption \cite{pan2021reconfigurable,you2020channel}. Due to the aforementioned attractive characteristics, IRS becomes a focal point of research in wireless communications.
\par The IRS-aided communications have been extensively investigated under various setups and objectives. By jointly optimizing active and passive beamforming, the IRS can significantly improve the communication performance in terms of spectral efficiency \cite{yu2019miso,zhou2020spectral,you2020energy}, sum rate \cite{pan2020intelligent,zhou2020intelligent,guo2020weighted,wang2019sum,8849960}, and energy efficiency \cite{huang2019reconfigurable,zeng2021energy,forouzanmehr2021energy,fang2020energy}. For example, the work \cite{yu2019miso} considered the spectral efficiency maximization problem via joint  active and passive beamforming. Two beamforming algorithms based on fixed point iteration and manifold optimization techniques were respectively proposed for solving this problem. Later on, considering the general multi-user case, the work \cite{pan2020intelligent} formulated a weighted sum rate (WSR) maximization problem, which is solved by exploiting the block coordinate descent (BCD) technique. Furthermore, taking the hardware limitation into consideration, the authors in \cite{zhou2020intelligent} proposed two beamforming algorithms with discrete phase shifts to maximize the sum rate, by invoking the majorization–minimization (MM) method. In addition to improving the spectral efficiency or the sum rate, IRS also has significant advantages in improving energy efficiency. For example, it was shown in \cite{huang2019reconfigurable} that through joint active and passive beamforming, the IRS-assisted MISO communication system can provide up to 300\% higher energy efficiency than the relay-assisted one.
\par In addition to using the IRS for communications, some works explored the potential of the IRS in wireless localization \cite{hu2018beyond,he2020large,he2020adaptive,elzanaty2021reconfigurable,wang2021joint,zhang2020towards,zhang2021metalocalization}. In \cite{hu2018beyond}, the authors first introduced the IRS into wireless localization and derived the Cramer-Rao lower bounds (CRLB) for positioning with IRS. It was proved that the distributed IRS system can achieve better average CRLB than the centralized IRS system. Then,  the work \cite{he2020large} investigated an IRS-assisted 2D mmWave positioning system and demonstrated that the positioning accuracy increases with the number of IRS elements. Later on, a more practical IRS-aided 3D mmWave localization system was studied in \cite{elzanaty2021reconfigurable}. By using the received signal strength (RSS) based positioning technique, an IRS-aided multi-user location sensing algorithm was proposed in \cite{zhang2021metalocalization}. It was shown that with the assistance of the IRS, the localization error is reduced by at least 3 times. Furthermore, through an angle of arrival (AoA) based iterative positioning algorithm, a centimeter-level positioning accuracy was achieved  in \cite{wang2021joint}.

\par All the aforementioned works considered the design of IRS-aided communications and localization, separately. However, it was demonstrated in \cite{wang2021joint2} that there exists a trade-off between communication and localization performance. From this perspective, the work \cite{wang2021joint} established an IRS-aided mmWave-MIMO based joint localization and communication system, where the time allocation ratio for localization is optimized to balance the performance of communication and localization. In this paper, instead of allocating different time slots for communication and localization like \cite{wang2021joint}, we propose to conduct localization and communication at the same time and frequency resources, thereby improving the spectral efficiency of the IRS-aided system, and  also propose to use the sensed location information for enhancing communication performance. Specifically, we propose an IRS-enabled multi-user integrated sensing and communication (ISAC) framework, where a distributed semi-passive IRS is deployed to enable  multi-user  concurrent localization and communication. The detailed working process of the IRS-enabled multi-user ISAC system is designed, including transmission protocol, multi-user location sensing, and beamforming design. The main contributions of this paper are summarized as follows.
\begin{itemize}
	\item We construct a novel 3D multi-user ISAC framework based on a distributed IRS architecture to realize simultaneous communication and localization. The whole transmission period is composed of the ISAC period with two time blocks and the pure communication (PC) period. During the ISAC period, the passive sub-IRS assists the uplink transmission between the base station (BS) and multiple users, and meanwhile two semi-passive sub-IRSs conduct multi-user location sensing by using the uplink communication signals sent by multiple users to the BS. The sensed users' locations in the first time block will be used for beamforming design in the second time block. During the PC period, all three sub-IRSs assist the uplink transmission by exploiting the users' locations sensed  in the second time block of the ISAC period for beamforming design.
	\item We propose a multi-user location sensing algorithm based on the uplink  communication signals  transmitted from multiple users to the BS, thereby removing the requirement of sending dedicated positioning reference signals.
	Specifically, we first estimate the effective AoA pairs corresponding to user-IRS links. Then, we estimate the path loss corresponding to each pair of effective AoAs. Finally, based on the estimated AoA pairs and their corresponding path losses, an AoA matching algorithm is proposed to determine users' locations. 
	\item We propose two novel beamforming algorithms with discrete phase shifts to maximize the sum rate for the ISAC period and the PC period, respectively, where the sensed location information is exploited for joint active and passive beamforming, thereby avoiding  high-overhead channel estimation.
	\item Simulation results show that the proposed multi-user location sensing algorithm achieves millimeter-level positioning accuracy, demonstrating the benefit of the proposed IRS-enabled ISAC framework in realizing high-accuracy localization.
	Moreover, the proposed sensing-based beamforming algorithms with sensed users' locations and discrete phase shifts achieve similar performance to the benchmark assuming perfect channel state information (CSI) and continuous phase shifts, validating the effectiveness of beamforming designs using sensed location information.
\end{itemize}
\par The remainder of this paper is organized as follows. Section~\ref{section2} introduces the system model of the IRS-aided multi-user ISAC system. Section~\ref{section3} presents a multi-user location sensing algorithm, while Section~\ref{section4} presents two sensing-based beamforming algorithms for the ISAC and PC periods, respectively. Numerical results are  provided in Section~\ref{section5}. Finally, Section~\ref{section6} concludes this paper.

\emph{{Notations:}} Vectors and matrices are denoted by boldface lower case and boldface upper case, respectively. The superscripts $\left( \cdot \right) ^T$, $\left( \cdot \right) ^H$ ,$\left( \cdot \right) ^{-1}$, $\left( \cdot \right) ^{\dagger}$, and $\left( \cdot \right) ^*$ denote  the operations of transpose, Hermitian transpose, inverse, pseudo-inverse and conjugate, respectively. The Euclidean norm, absolute value, Kronecker product are respectively denoted by $\left\| \cdot \right\|$, $\left| \cdot \right|$ and $\otimes $. $\mathbb{E} \left\{ \cdot \right\} $ denotes the statistical expectation. Moreover, $\mathcal{C} \mathcal{N} ( 0,\sigma ^2 ) $ denotes the circularly symmetric complex Gaussian (CSCG) distribution with zero mean  and variance $\sigma ^2$. For matrices,
$\left[ \cdot \right] _{ij}$ denotes the $(i,j)$-th element of a matrix, $\text{tr}\left( \cdot \right) $ represents the matrix trace, $\text{diag}\left( \cdot \right) $ denotes a square diagonal matrix with the elements in $\left( \cdot \right) $ on its main diagonal, and $\mathbf{1}_{N\times M}$ denotes an $N\!\times\! M$ all-one matrix. For vectors, $\left[ \cdot \right] _i$ denotes the $i$-th entry of a vector. Besides, $j$ in $e^{j\theta}$ denotes the imaginary unit. 

\section{System Model}\label{section2}

\begin{figure}[htbp]
  \centering
  \includegraphics[width=3.2in]{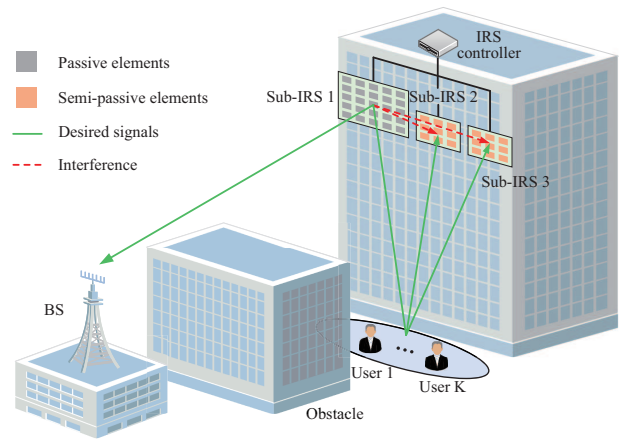}
  \caption{An IRS-enabled multi-user ISAC system.}
  \label{system_model1}
\end{figure}

As illustrated in Fig.~\ref{system_model1}, we consider an IRS-aided multi-user uplink communication system operating in the mmWave band, where a distributed semi-passive IRS with $M$ reflecting elements is deployed to assist the uplink data transmission from $K$ single-antenna users to a multi-antenna BS. We consider that the line-of-sight (LoS) paths between the BS and users are obstructed, and the IRS is deployed to establish strong virtual line-of-sight (VLoS) reflection paths between them. The distributed semi-passive IRS is composed of 3 sub-IRSs. The first sub-IRS is passive and consists of $M_1$ passive reflecting elements, while the $i$-th ($i$ = 2, 3) sub-IRS is semi-passive and consists of $M_i$ ($M_i \ll M_1$) semi-passive reflecting elements, which are capable of both sensing and reflecting. For ease of practical implementation, the phase shift of IRS takes values from a finite set $\mathcal{F} =\{ 0,\frac{2\pi}{2^b},\cdots ,\frac{2\pi}{2^b}( 2^b-1 ) \} $, where $b$ is the bit-quantization number. The BS has an $N$-element uniform linear array (ULA) along the y axis, while the $i$-th sub-IRS has an $M_{y,i} \times M_{z,i}$ uniform rectangular array (URA) lying on the $y$-$o$-$z$ plane. Moreover, there is a backhaul link connecting the BS and the IRS for exchanging information. 
In this paper, both user-IRS and IRS-BS channels are modeled as  quasi-static block-fading channels, which remain almost unchanged during each coherence block but vary from one block to another.
\vspace{-2mm}
\subsection{ISAC Transmission Protocol}
\begin{figure}[htbp]
  \centering
  \includegraphics[width=3.2in]{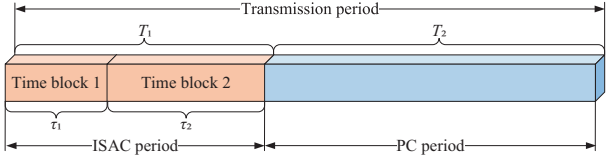}
  \caption{Illustration of the ISAC transmission protocol.}
  \label{system_model2}
\end{figure}
As illustrated in Fig.~\ref{system_model2}, we consider the transmission period  within one coherence block, which is composed of the ISAC period with $T_1$ time slots and the PC period with $T_2$ time slots. The ISAC period is divided into time block 1 with $\tau_1$ time slots and time block 2 with $\tau_2$ time slots. During the ISAC period, the passive sub-IRS assists the uplink data transmission by reflecting, and meanwhile the two semi-passive sub-IRSs estimate users' locations based on the communication signals transmitted from multiple users to the BS. Specifically, in the first time block, the passive sub-IRS generates a random phase shift beam due to the unavailability of CSI. In the second time block, the phase shift beam of the passive sub-IRS is properly designed to assist uplink data transmission by utilizing the users' locations obtained in the first time block. During the PC period, all three sub-IRSs operate in the reflecting  mode to assist the uplink data transmission. And  joint active and passive beamforming is designed, based on more accurate users' locations estimated in the second time block of the ISAC period.
\vspace{-2mm}
\subsection{Signal Model}
\subsubsection{ISAC Period}
In the ISAC period, only the passive sub-IRS (i.e., the first sub-IRS) operates in the reflecting mode to assist uplink transmission. During the $n$-th time block, the $k$-th user sends $\sqrt{\rho}s_k\left( t \right)$, satisfying $\left| s_k\left( t \right) \right|$ = 1, to the BS at time slot $t\in \mathcal{N} _n=\{ \left( n-1 \right) \tau _1+1,\cdots ,\tau _1+\left( n-1 \right) \tau _2 \} $, where $\rho$ denotes the transmit power. The received signal at the BS is given by
\begin{align} \label{SMA1}
y\left( t \right) =&\sqrt{\rho}\sum_{k=1}^K{\left[ \mathbf{w}_{k}^{\left( n \right)} \right] ^H\mathbf{H}_{\mathrm{I}2\mathrm{B},1}\mathbf{\Theta }_{1}^{\left( n \right)}\mathbf{h}_{\mathrm{U}2\mathrm{I},1,k}\,\,s_k\left( t \right)}\notag\\
&+\sum_{k=1}^K{\left[ \mathbf{w}_{k}^{\left( n \right)} \right] ^H\mathbf{n}_{\mathrm{BS}}\left( t \right)},\,t\in \mathcal{N} _n,\,n=1,2,
\end{align}
where $\mathbf{w}_{k}^{\left( n \right)}$, satisfying $\| \mathbf{w}_{k}^{\left( n \right)} \| =1$, represents the BS combining vector of the $k$-th user in the $n$-th time block, $\mathbf{H}_{\mathrm{I}2\mathrm{B},i}\in \mathbb{C} ^{N\times M_i}$ and $\mathbf{h}_{\mathrm{U}2\mathrm{I},i,k}\in \mathbb{C} ^{M_i\times 1}$ denote the baseband equivalent channels from the $i$-th sub-IRS to the BS and from the $k$-th user to the $i$-th sub-IRS, respectively. The phase shift matrix of the first sub-IRS in the $n$-th time block is defined as $\mathbf{\Theta }_{1}^{\left( n \right)}=\mathrm{diag}( \boldsymbol{\xi }_{1}^{\left( n \right)} ) $, with the phase shift beam being $\boldsymbol{\xi }_{1}^{\left( n \right)}=[ e^{j\vartheta _{1,1}^{\left( n \right)}},\cdots ,e^{j\vartheta _{1,m}^{\left( n \right)}},\cdots ,e^{j\vartheta _{1,M_1}^{\left( n \right)}} ] ^T$. In addition, $\mathbf{n}_{\mathrm{BS}}$ represents the additive white Gaussian noise (AWGN) at the BS, whose elements follow the complex Gaussian distribution $\mathcal{C} \mathcal{N} ( 0,\sigma _{0}^{2} )$.
\par The two semi-passive sub-IRSs operate in the sensing mode, and the received signal at the $i$-th sub-IRS is given by 
\begin{align} \label{SMA2}
\mathbf{x}_i\!\left( t \right) \!=\!&\sqrt{\rho}\sum_{k=1}^K{\!\mathbf{h}_{\mathrm{U}2\mathrm{I},i,k}\,s_k\!\left( t \right)}\!+\!\sqrt{\rho}\mathbf{H}_{\mathrm{I}2\mathrm{I},i}\mathbf{\Theta }_{1}^{\left( n \right)}\!\sum_{k=1}^K{\!\mathbf{h}_{\mathrm{U}2\mathrm{I},1,k}\,s_k\!\left( t \right)}\notag
\\
&+\mathbf{n}_i\left( t \right) ,\,i=2,3,\,t\in \mathcal{N} _n,\,n=1,2,
\end{align}
where $\mathbf{H}_{\mathrm{I}2\mathrm{I},i}\in \mathbb{C} ^{M_i\times M_1}$ denotes the baseband equivalent channel from the passive sub-IRS to the $i$-th sub-IRS, and $\mathbf{n}_i$ represents the AWGN at the $i$-th sub-IRS, whose elements follow the complex Gaussian distribution $\mathcal{C} \mathcal{N} ( 0,\sigma _{0}^{2} ) $.
\par The instantaneous achievable rate of the $k$-th user during the ISAC period is given by
\begin{align} \label{SMA3}
&R_k\left( t \right)=\\
&\log _2\!\left( 1\!+\!\frac{\rho \left| \left[ \mathbf{w}_{k}^{\left( n \right)} \right] ^H\mathbf{H}_{\mathrm{I}2\mathrm{B},1}\mathbf{\Theta }_{1}^{\left( n \right)}\mathbf{h}_{\mathrm{U}2\mathrm{I},1,k} \right|^2}{\rho \sum\nolimits_{j\ne k}^K{\left| \left[ \mathbf{w}_{k}^{\left( n \right)} \right] ^H\mathbf{H}_{\mathrm{I}2\mathrm{B},1}\mathbf{\Theta }_{1}^{\left( n \right)}\mathbf{h}_{\mathrm{U}2\mathrm{I},1,j} \right|^2}\!+\!\sigma _{0}^{2}} \right),\notag\\
&\qquad\qquad\qquad\qquad\qquad t\in \mathcal{N} _n,\,n=1,2.\notag
\end{align}

\vspace{-0.2cm}
\subsubsection{PC Period}
In the PC period, all the three sub-IRSs operate in the reflecting mode to assist uplink transmission. The $k$-th user sends $\sqrt{\rho}s_k\left( t \right) $ to the BS at time slot  $t\in \mathcal{T} _2 \triangleq \left\{ T_1+1,\cdots,T_1+T_2 \right\} $. The received signal at the BS is
\begin{align} \label{SMA4}
y\left( t \right) =&\sqrt{\rho}\sum_{k=1}^K{\mathbf{w}_{k}^{H}\left( t \right) \mathbf{H}_{\mathrm{I}2\mathrm{B}}\mathbf{\Theta }\left( t \right) \mathbf{h}_{\mathrm{U}2\mathrm{I},k}\,\,s_k\left( t \right)}\notag\\
&+\sum_{k=1}^K{\mathbf{w}_{k}^{H}\left( t \right) \mathbf{n}_{\mathrm{BS}}\left( t \right)},\,t\in \mathcal{T} _2,
\end{align}
where $\mathbf{H}_{\mathrm{I}2\mathrm{B}}\!\triangleq\! [ \mathbf{H}_{\mathrm{I}2\mathrm{B},1},\mathbf{H}_{\mathrm{I}2\mathrm{B},2},\mathbf{H}_{\mathrm{I}2\mathrm{B},3} ] \!\in\! \mathbb{C} ^{N\times M}$ and $\mathbf{h}_{\mathrm{U}2\mathrm{I},k}$ $\triangleq[ \mathbf{h}_{\mathrm{U}2\mathrm{I},1,k}^{T},\mathbf{h}_{\mathrm{U}2\mathrm{I},2,k}^{T},\mathbf{h}_{\mathrm{U}2\mathrm{I},3,k}^{T} ] ^T\!\in\! \mathbb{C} ^{M\times 1}$ denote the baseband equivalent channels from the IRS to the BS and from the $k$-th user to the IRS, respectively. The phase shift matrix of the whole IRS is defined as $\mathbf{\Theta }=\mathrm{diag}( \boldsymbol{\xi } )$, where $\boldsymbol{\xi }\triangleq [ \boldsymbol{\xi }_{1}^{T},\boldsymbol{\xi }_{2}^{T},\boldsymbol{\xi }_{3}^{T} ] ^T$  with the phase shift beam being $\boldsymbol{\xi }_i=[ e^{j\vartheta _{i,1}},\cdots ,e^{j\vartheta _{i,m}},\cdots ,e^{j\vartheta _{i,M_i}} ] ^T$.
\par The instantaneous achievable rate of the $k$-th user during the PC period is given by
\begin{align} \label{SMA5}
&R_k\!\left( t \right) =
\\
&\log _2\!\left( 1\!+\!\frac{\rho \left| \mathbf{w}_{k}^{H}\left( t \right) \mathbf{H}_{\mathrm{I}2\mathrm{B}}\mathbf{\Theta }\left( t \right) \mathbf{h}_{\mathrm{U}2\mathrm{I},k} \right|^2}{\rho \!\sum\nolimits_{j\ne k}^K{\left| \mathbf{w}_{k}^{H}\left( t \right) \mathbf{H}_{\mathrm{I}2\mathrm{B}}\mathbf{\Theta }\left( t \right) \mathbf{h}_{\mathrm{U}2\mathrm{I},j} \right|^2}\!\!+\!\sigma _{0}^{2}} \right) ,t\in \mathcal{T} _2.\notag
\end{align}
\subsection{Channel Model}
In general, the IRS operating in the mmWave band is deployed with LoS paths to both the BS and users. Hence, the channel from the $i$-th\! sub-IRS \!to\! the BS \!is\! modelled as \!\!\cite{9235486}
\begin{align} \label{SMB1}
\mathbf{H}_{\mathrm{I}2\mathrm{B},i}\!=\!\alpha _{\mathrm{I}2\mathrm{B},i}\mathbf{a}\left( u_{\mathrm{I}2\mathrm{B},i}^{\mathrm{A}} \right) \mathbf{b}_{i}^{H}\!\left( u_{\mathrm{I}2\mathrm{B},i}^{\mathrm{D}},v_{\mathrm{I}2\mathrm{B},i}^{\mathrm{D}} \right),i=1,2,3,
\end{align}
where $\alpha _{\mathrm{I}2\mathrm{B},i}$ denotes the complex channel gain for the link from the $i$-th sub-IRS to the BS, $\mathbf{a}$ and $\mathbf{b}_i$ are the array response vectors for the BS and the $i$-th sub-IRS, respectively. The two effective angles of departure (AoDs) $u_{\mathrm{I}2\mathrm{B},i}^{\mathrm{D}}$ and $v_{\mathrm{I}2\mathrm{B},i}^{\mathrm{D}}$ as well as the effective AoA $u_{\mathrm{I}2\mathrm{B},i}^{\mathrm{A}}$ are respectively defined as
\begin{align} 
&u_{\mathrm{I}2\mathrm{B},i}^{\mathrm{D}}=2\pi \frac{d_{\mathrm{IRS}}}{\lambda}\cos \left( \gamma _{\mathrm{I}2\mathrm{B},i}^{\mathrm{D}} \right) \sin \left( \varphi _{\mathrm{I}2\mathrm{B},i}^{\mathrm{D}} \right) ,\\
&v_{\mathrm{I}2\mathrm{B},i}^{\mathrm{D}}=2\pi \frac{d_{\mathrm{IRS}}}{\lambda}\sin \left( \gamma _{\mathrm{I}2\mathrm{B},i}^{\mathrm{D}} \right),\\ &u_{\mathrm{I}2\mathrm{B},i}^{\mathrm{A}}=2\pi \frac{d_{\mathrm{BS}}}{\lambda}\sin \left( \theta _{\mathrm{I}2\mathrm{B},i}^{\mathrm{A}} \right),
\end{align}
where $\lambda $ denotes the carrier wavelength, $d_{\mathrm{IRS}}$ and $d_{\mathrm{BS}}$ represent the distances between two adjacent reflecting elements of the IRS and two adjacent antennas of the BS, respectively. In addition,  $\theta _{\mathrm{I}2\mathrm{B},i}^{\mathrm{A}}$ denotes the AoA at the BS, $\gamma _{\mathrm{I}2\mathrm{B},i}^{\mathrm{D}}$ and $\varphi _{\mathrm{I}2\mathrm{B},i}^{\mathrm{D}}$ are the elevation and azimuth AoDs for the link from the $i$-th sub-IRS to the BS, respectively.
\par Similarly, the channel from the $k$-th user to the $i$-th sub-IRS is modelled as
\begin{align} \label{SMB2}
\mathbf{h}_{\mathrm{U}2\mathrm{I},i,k}= \,&\alpha _{\mathrm{U}2\mathrm{I},i,k}\mathbf{b}_i\left( u_{\mathrm{U}2\mathrm{I},i,k}^{\mathrm{A}},v_{\mathrm{U}2\mathrm{I},i,k}^{\mathrm{A}} \right),\notag\\
&i=1,2,3,\, k=1,\cdots,K,
\end{align}
where $\alpha _{\mathrm{U}2\mathrm{I},i,k}$ denotes the complex channel gain for the link from the $k$-th user to the $i$-th sub-IRS, and the two effective AoAs from  the $k$-th user to the $i$-th sub-IRS are defined as
\begin{align} 
&u_{\mathrm{U}2\mathrm{I},i,k}^{\mathrm{A}}=2\pi \frac{d_{\mathrm{IRS}}}{\lambda}\cos \left( \gamma _{\mathrm{U}2\mathrm{I},i,k}^{\mathrm{A}} \right) \sin \left( \varphi _{\mathrm{U}2\mathrm{I},i,k}^{\mathrm{A}} \right), 
\\
&v_{\mathrm{U}2\mathrm{I},i,k}^{\mathrm{A}}=2\pi \frac{d_{\mathrm{IRS}}}{\lambda}\sin \left( \gamma _{\mathrm{U}2\mathrm{I},i,k}^{\mathrm{A}} \right), 
\end{align}
where $\gamma_{\mathrm{U}2\mathrm{I},i,k}^{\mathrm{A}}$ and $\varphi_{\mathrm{U}2\mathrm{I},i,k}^{\mathrm{A}}$ denote the elevation and azimuth AoAs for the link from the $k$-th user to the $i$-th sub-IRS, respectively. 
\par Also, the channel from the passive sub-IRS to the $i$-th sub-IRS is modelled as
\begin{align} \label{SMB3}
\mathbf{H}_{\mathrm{I}2\mathrm{I},i}\!=\!\alpha _{\mathrm{I}2\mathrm{I},i}\mathbf{b}_i\!\left( u_{\mathrm{I}2\mathrm{I},i}^{\mathrm{A}},v_{\mathrm{I}2\mathrm{I},i}^{\mathrm{A}} \right) \!\mathbf{b}_{1}^{H}\!\!\left( u_{\mathrm{I}2\mathrm{I},i}^{\mathrm{D}},v_{\mathrm{I}2\mathrm{I},i}^{\mathrm{D}} \right)\!, i\!=\!2,3,
\end{align}
where $\alpha _{\mathrm{I}2\mathrm{I},i}$ denotes the complex channel gain for the link from the passive sub-IRS to the $i$-th sub-IRS. In addition, the two effective AoAs $u_{\mathrm{I}2\mathrm{I},i}^{\mathrm{A}}$ and $v_{\mathrm{I}2\mathrm{I},i}^{\mathrm{A}}$ as well as the two effective AoDs $u_{\mathrm{I}2\mathrm{I},i}^{\mathrm{D}}$ and $v_{\mathrm{I}2\mathrm{I},i}^{\mathrm{D}}$ are respectively defined as
\begin{align} 
&u_{\mathrm{I}2\mathrm{I},i}^{\mathrm{A}}=2\pi \frac{d_{\mathrm{IRS}}}{\lambda}\cos \left( \gamma _{\mathrm{I}2\mathrm{I},i}^{\mathrm{A}} \right) \sin \left( \varphi _{\mathrm{I}2\mathrm{I},i}^{\mathrm{A}} \right) ,
\\
&v_{\mathrm{I}2\mathrm{I},i}^{\mathrm{A}}=2\pi \frac{d_{\mathrm{IRS}}}{\lambda}\sin \left( \gamma _{\mathrm{I}2\mathrm{I},i}^{\mathrm{A}} \right),
\\
&u_{\mathrm{I}2\mathrm{I},i}^{\mathrm{D}}=2\pi \frac{d_{\mathrm{IRS}}}{\lambda}\cos \left( \gamma _{\mathrm{I}2\mathrm{I},i}^{\mathrm{D}} \right) \sin \left( \varphi _{\mathrm{I}2\mathrm{I},i}^{\mathrm{D}} \right) ,
\\
&v_{\mathrm{I}2\mathrm{I},i}^{\mathrm{D}}=2\pi \frac{d_{\mathrm{IRS}}}{\lambda}\sin \left( \gamma _{\mathrm{I}2\mathrm{I},i}^{\mathrm{D}} \right),
\end{align}
where $\gamma _{\mathrm{I}2\mathrm{I},i}^{\mathrm{A}}$/ $\varphi _{\mathrm{I}2\mathrm{I},i}^{\mathrm{A}}$ denotes the elevation/azimuth AoA and  $\gamma _{\mathrm{I}2\mathrm{I},i}^{\mathrm{D}}$/$\varphi _{\mathrm{I}2\mathrm{I},i}^{\mathrm{D}}$ denotes the elevation/azimuth AoD from the passive sub-IRS to the $i$-th sub-IRS.
\par Furthermore, we consider that $d_{\mathrm{BS}}=d_{\mathrm{IRS}}=\frac{\lambda}{2}$. As such, the array response vectors for the BS and the $i$-th sub-IRS are  respectively given by
\begin{align} 
\mathbf{a}\left( u \right) &=\left[ 1,\cdots,e^{j\left( n-1 \right) u},\cdots,e^{j\left( N-1 \right) u} \right] ^T,\\
\mathbf{b}_i\left( u,v \right) &=\left[ 1,\cdots ,e^{j\left( n-1 \right) u},\cdots ,e^{j\left( M_{y,i}-1 \right) u} \right] ^T\notag
\\
&\quad\,\,\otimes \left[ 1,\cdots ,e^{j\left( m-1 \right) v},\cdots ,e^{j\left( M_{z,i}-1 \right) v} \right] ^T.
\end{align}

\section{multi-user location sensing}\label{section3}
In this section, we design a multi-user location sensing algorithm  based on the received communication signals at the two semi-passive sub-IRSs. Specifically, we first estimate the effective AoA pairs at the two semi-passive IRSs,  by combining the use of  the  forward-backward spatial smoothing (FBSS) technique, the total least squares-estimation of signal parameters via rotational invariance technique (TLS-ESPRIT), and the multiple signal classification (MUSIC) technique. After excluding the AoA pairs corresponding to the links from the passive sub-IRS to the two semi-passive sub-IRSs, we obtain the effective AoA pairs corresponding to the links from multiple users to the semi-passive sub-IRSs. Then, we estimate the path losses corresponding to these effective AoA pairs. Finally, based on the estimated effective AoA pairs and their corresponding path losses, an AoA matching algorithm is proposed to determine users' locations.
\subsection{Estimate Effective AoAs}
To reduce the coherency of the received signals at the semi-passive sub-IRS, we first adopt the FBSS technique to preprocess the received signals. Then, we separately estimate the effective AoAs corresponding to the $y$ axis (i.e., $u_{\mathrm{U}2\mathrm{I},i,k}^{\mathrm{A}}$ and $u_{\mathrm{I}2\mathrm{I},i}^{\mathrm{A}}$) and the $z$ axis (i.e., $v_{\mathrm{U}2\mathrm{I},i,k}^{\mathrm{A}}$ and $v_{\mathrm{I}2\mathrm{I},i}^{\mathrm{A}}$), by  invoking the TLS-ESPRIT algorithm.
Finally, we pair the effective AoAs corresponding to the $y$ axis with those corresponding to the $z$ axis by invoking the MUSIC  algorithm.
\subsubsection{Preprocess the received signals by using the FBSS technique}
\par
\begin{figure}[htbp]
  \centering
  \includegraphics[width=3.2in]{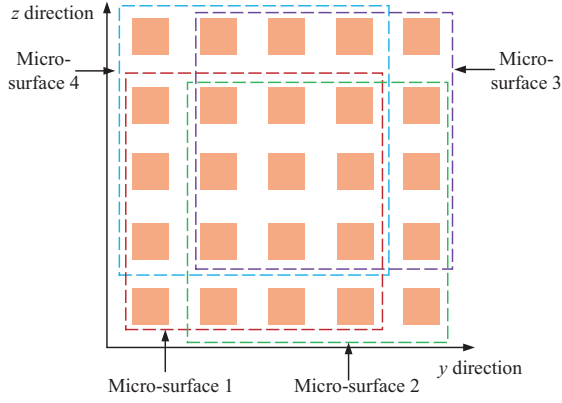}
  \caption{An example of the micro-surface.}
  \label{location sensing1}
  \vspace{-0.3cm}
\end{figure}

 Without loss of generality, we focus on the estimation of the effective AoAs at the $i$-th sub-IRS in the $n$-th time block ($i\in \left\{ 2,3 \right\}$ and $n\in \left\{ 1,2 \right\}$). We construct a set of $N_{\mathrm{micro},i}$ micro-surfaces for the $i$-th sub-IRS. Each micro-surface is composed of  $L_{\mathrm{micro},i}=Q_{y,i}\times Q_{z,i}$ semi-passive elements and shifted by one row along the $z$ direction or one column along the $y$ direction from the preceding micro-surface. For example, we construct a set of 4 micro-surfaces each with $4\times4$ semi-passive elements, as illustrated in Fig.~\ref{location sensing1}.

Denote the received signals at the $m$-th micro-surface of the $i$-th sub-IRS during the $n$-th time block by $\mathbf{x}_{i,m}\left( t \right) \in \mathbb{C} ^{L_{\mathrm{micro},i}\times 1},t \in \mathcal{N}_n$, and define their auto-correlation matrix as $\mathbf{R}_{\mathrm{micro},i}^{\left( n \right)}\triangleq \mathbb{E} \{ \mathbf{x}_{i,m}( t ) [ \mathbf{x}_{i,m}( t ) ] ^H \} ,t\in \mathcal{N} _n$.
According to the FBSS technique, we estimate $\mathbf{R}_{\mathrm{micro},i}^{\left( n \right)}$ as
\begin{align}
\mathbf{\hat{R}}_{\mathrm{micro},i}^{\left( n \right)}=&\frac{1}{2\tau _nN_{\mathrm{micro},i}}\!\sum_{t\in \mathcal{N} _n}^{}{\!\!\!\sum_{m=1}^{N_{\mathrm{micro},i}}{\!\!\!\left\{ \mathbf{x}_{i,m}\left( t \right) \left[ \mathbf{x}_{i,m}\left( t \right) \right] ^H \right.}}\notag
\\
&\qquad\qquad\quad+\left. \mathbf{J}\left[ \mathbf{x}_{i,m}\left( t \right) \right] ^*\left[ \mathbf{x}_{i,m}\left( t \right) \right] ^T\mathbf{J} \right\} ,
\end{align}
where $\mathbf{J}$ denotes the exchange matrix, with the 1 elements residing on its counterdiagonal and all other elements being zero. Then, we perform eigenvalue decomposition of $\mathbf{\hat{R}}_{\mathrm{micro},i}^{\left( n \right)}$
\begin{align} 
\mathbf{\hat{R}}_{\mathrm{micro},i}^{\left( n \right)}=\mathbf{U}_{i}^{\left( n \right)}\mathrm{diag}\left( \lambda _{i,1}^{\left( n \right)},\cdots,\lambda _{i,L_{\mathrm{micro},i}}^{\left( n \right)} \right) \left[ \mathbf{U}_{i}^{\left( n \right)} \right] ^H,
\end{align}
where $\mathbf{U}_{i}^{\left( n \right)}\!\triangleq\! [ \mathbf{u}_{i,1}^{\left( n \right)},\cdots,\mathbf{u}_{i,L_{\mathrm{micro},i}}^{\left( n \right)} ] $ and the eigenvalues $\lambda _{i,1}^{\left( n \right)},$ $\cdots,\lambda _{i,L_{\mathrm{micro},i}}^{\left( n \right)}$ are in descending order.
\vspace{-0.4cm}
\subsubsection{Estimate $u_{\mathrm{U}2\mathrm{I},i,k}^{\mathrm{A}}$ and $u_{\mathrm{I}2\mathrm{I},i}^{\mathrm{A}}$ by using the TLS ESPRIT algorithm}
\begin{figure}[htbp]
  \centering
  \subfigure[]
  {  
  \label{example.sub.1}
  \includegraphics[width=1.65in]{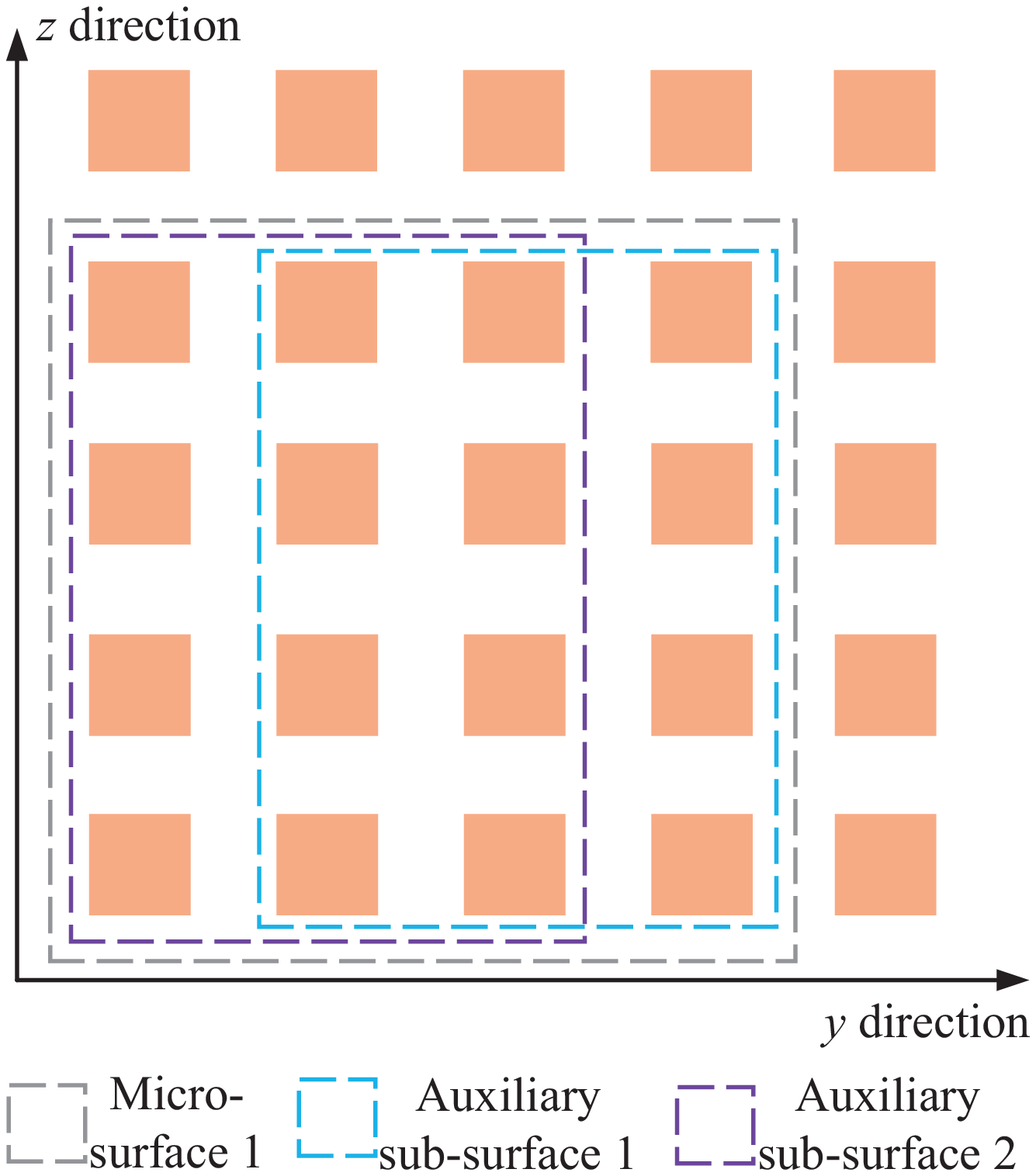}}~\subfigure[]
  {
  \label{example.sub.2}
  \includegraphics[width=1.65in]{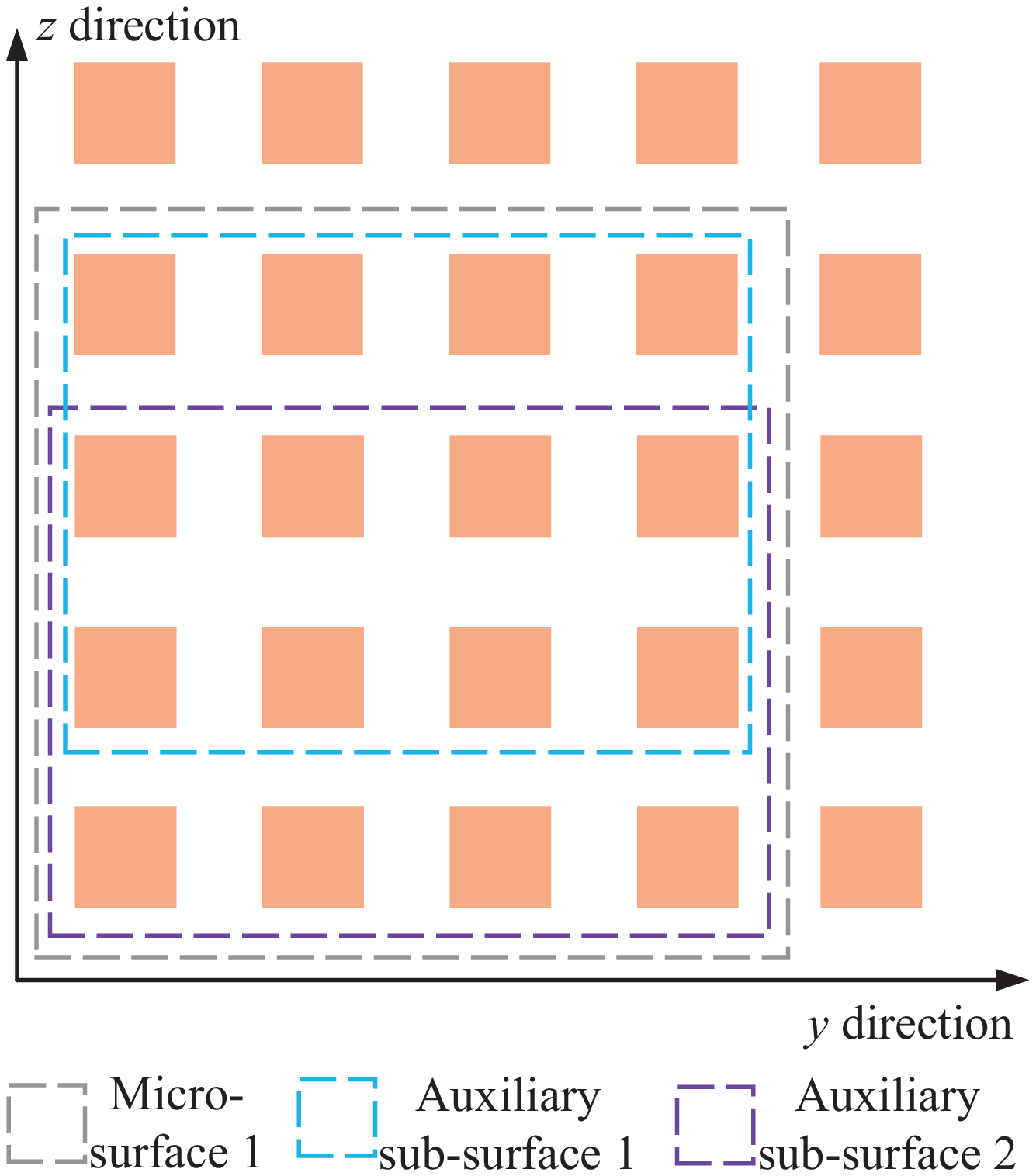}
  }
  \caption{An example of the auxiliary sub-surface.}
  \label{example}
\end{figure}
As shown in Fig.~\ref{example.sub.1}, for the first micro-surface of the $i$-th sub-IRS, we construct two auxiliary sub-surfaces each with $L_{\mathrm{aux},i}=( Q_{y,i}-1 ) \times Q_{z,i}$ semi-passive elements, satisfying $L_{\mathrm{aux},i}\geqslant K\!+\!2$. The signal sub-space corresponding to the two auxiliary sub-surfaces is given by
\begin{align}
\mathbf{U}_{\mathrm{S},im}^{\left( n \right)}\triangleq \mathbf{J}_m\mathbf{U}_{\mathrm{S},i}^{\left( n \right)}\,\,,m=1,2,
\end{align}
where $\mathbf{U}_{\mathrm{S},i}^{\left( n \right)}\triangleq [ \mathbf{u}_{i,1}^{\left( n \right)},\cdots,\mathbf{u}_{i,K+1}^{\left( n \right)} ] \!\in \!\mathbb{C} ^{L_{\mathrm{micro},i}\times (K+1)}$, $\mathbf{J}_m\!\in\! \mathbb{R} ^{L_{\mathrm{aux},i}\times L_{\mathrm{micro},i}}$ is a selecting matrix whose elements are either 1 or 0. If the $j$-th reflecting element of the micro-surface 1 is selected as the $i$-th element of the auxiliary sub-surface $m$, then we set $\left[ \mathbf{J}_m \right] _{ij}=1$. Otherwise, we set $\left[ \mathbf{J}_m \right] _{ij}=0$.
\par Then, we define $\mathbf{C}_{i}^{\left( n \right)}\triangleq [ \mathbf{U}_{\mathrm{S},i1}^{\left( n \right)},\mathbf{U}_{\mathrm{S},i2}^{\left( n \right)} ] ^H[ \mathbf{U}_{\mathrm{S},i1}^{\left( n \right)},\mathbf{U}_{\mathrm{S},i2}^{\left( n \right)} ] \in \mathbb{C} ^{\left( 2K+2 \right) \times \left( 2K+2 \right)}$ and perform eigendecomposition of it, which yields
\begin{align}
    \mathbf{C}_{i}^{\left( n \right)}=\left[ \begin{matrix}
	\mathbf{V}_{i,11}^{\left( n \right)}&		\mathbf{V}_{i,12}^{\left( n \right)}\\
	\mathbf{V}_{i,21}^{\left( n \right)}&		\mathbf{V}_{i,22}^{\left( n \right)}\\
\end{matrix} \right] \mathbf{\Lambda }_{C,i}^{\left( n \right)}\left[ \begin{matrix}
	\mathbf{V}_{i,11}^{\left( n \right)}&		\mathbf{V}_{i,12}^{\left( n \right)}\\
	\mathbf{V}_{i,21}^{\left( n \right)}&		\mathbf{V}_{i,22}^{\left( n \right)}\\
\end{matrix} \right] ^H,
\end{align}
where $\mathbf{\Lambda }_{C,i}^{\left( n \right)}\triangleq \mathrm{diag}( \lambda _{\mathrm{C},i1}^{\left( n \right)},\cdots,\lambda _{\mathrm{C},i(2K+2)}^{\left( n \right)} ) $ with its eigenvalues in descending order, $\mathbf{V}_{i,12}^{\left( n \right)}$ and $\mathbf{V}_{i,22}^{\left( n \right)}$ are two $( K\!+\!1 ) \times ( K\!+\!1 ) $ matrices. 

After obtaining $\mathbf{V}_{i,12}^{\left( n \right)}$ and $\mathbf{V}_{i,22}^{\left( n \right)}$, we calculate 
\begin{align}
    \mathbf{\Phi }_{\mathrm{TLS},i}^{\left( n \right)}=-\mathbf{V}_{i,12}^{\left( n \right)}\left[ \mathbf{V}_{i,22}^{\left( n \right)} \right] ^{-1}.
\end{align}

By performing eigenvalue decomposition of $\mathbf{\Phi }_{\mathrm{TLS},i}^{\left( n \right)}$, we obtain its eigenvalues as $\lambda _{\mathrm{TLS},il}^{\left( n \right)}, l=1,\cdots ,K\!+\!1$.
\par Finally, we estimate the effective AoAs at the $i$-th sub-IRS corresponding to the $y$ axis as 
\begin{align}
\check{u}_{il}^{\left( n \right)}=\mathrm{angle}\left( \lambda _{\mathrm{TLS},il}^{\left( n \right)} \right) ,l=1,\cdots,K\!+\!1.
\end{align}
The estimators of $u_{\mathrm{U}2\mathrm{I},i,k}^{\mathrm{A}}$ and $u_{\mathrm{I}2\mathrm{I},i}^{\mathrm{A}}$ in the $n$-th time block belong to $\,\mathcal{U} _{i}^{\left( n \right)}\triangleq \{ \check{u}_{il}^{\left( n \right)},l\!=\!1,\cdots,K\!+\!1 \} $, i.e., $\hat{u}_{\mathrm{U}2\mathrm{I},i,k}^{\mathrm{A},\left( n \right)},\hat{u}_{\mathrm{I}2\mathrm{I},i}^{\mathrm{A},\left( n \right)}\in \mathcal{U} _{i}^{\left( n \right)}$.
\subsubsection{Estimate $v_{\mathrm{U}2\mathrm{I},i,k}^{\mathrm{A}}$ and $v_{\mathrm{I}2\mathrm{I},i}^{\mathrm{A}}$ by using the TLS-ESPRIT algorithm}
As shown in Fig.~\ref{example.sub.2}, for the first micro-surface of the $i$-th sub-IRS, we construct two auxiliary sub-surfaces each with $\tilde{L}_{\mathrm{aux},i}=Q_{y,i}\times \left( Q_{z,i}-1 \right)$ semi-passive elements, satisfying $\tilde{L}_{\mathrm{aux},i}\geqslant K\!+\!2$. Following the similar process of estimating $u_{\mathrm{U}2\mathrm{I},i,k}^{\mathrm{A}}$ and $u_{\mathrm{I}2\mathrm{I},i}^{\mathrm{A}}$, we estimate the effective AoAs at the $i$-th sub-IRS corresponding to the $z$ axis as $\check{v}_{il}^{\left( n \right)}, l=1,\cdots ,K\!+\!1$. The estimators of $v_{\mathrm{U}2\mathrm{I},i,k}^{\mathrm{A}}$ and $v_{\mathrm{I}2\mathrm{I},i}^{\mathrm{A}}$ in the $n$-th time block belong to $\mathcal{V} _{i}^{\left( n \right)}\triangleq \{ \check{v}_{il}^{\left( n \right)},l=1,\cdots,K\!+\!1 \} $, i.e., $\hat{v}_{\mathrm{U}2\mathrm{I},i,k}^{\mathrm{A},\left( n \right)},\hat{v}_{\mathrm{I}2\mathrm{I},i}^{\mathrm{A},\left( n \right)}\in \mathcal{V} _{i}^{\left( n \right)}$.
\subsubsection{Pair $\check{u}_{il}^{\left( n \right)}$ and $\check{v}_{il}^{\left( n \right)}$ by using the MUSIC algorithm}
Let
\begin{align}
\check{f}_{i,ls}^{\left( n \right)}\triangleq \mathbf{b}_{\mathrm{micro},i}^{H}\left( \check{u}_{il}^{\left( n \right)},\check{v}_{is}^{\left( n \right)} \right) \mathbf{U}_{\mathrm{N},i}^{\left( n \right)},
\end{align}
where $\!\mathbf{U}_{\mathrm{N},i}^{\left( n \right)}\!\triangleq\! [ \mathbf{u}_{i,\left( \!K\!+2 \right)}^{\left( n \right)}\!,\!\cdots\! ,\mathbf{u}_{i,L_{\mathrm{micro},i}}^{\left( n \right)}] \!\!\in \!\mathbb{C} ^{L_{\mathrm{micro},i}\!\times \!\left( L_{\mathrm{micro},i}\!-\!K\!-\!1 \right)}$, 
and $\mathbf{b}_{\mathrm{micro},i}$ denotes the array response vector of the micro-surface on the $i$-th sub-IRS. Then, we compute
\begin{align} 
f\left( \check{u}_{il}^{\left( n \right)},\check{v}_{is}^{\left( n \right)} \right) =\check{f}_{i,ls}^{\left( n \right)}\left[ \check{f}_{i,ls}^{\left( n \right)} \right] ^H,l,s=1,\cdots,K\!+\!1,
\end{align}
and choose the $K\!+\!1$ minima $f( \hat{u}_{il}^{\left( n \right)},\hat{v}_{il}^{\left( n \right)} )$, $l=1,\cdots ,K\!+\!1$, where $\hat{u}_{il}^{\left( n \right)}\in \mathcal{U} _{i}^{\left( n \right)}$ and $\hat{v}_{il}^{\left( n \right)}\in \mathcal{V} _{i}^{\left( n \right)}$. As such, we obtain $K\!+\!1$ pairs of effective AoAs $( \hat{u}_{il}^{\left( n \right)},\hat{v}_{il}^{\left( n \right)} ),l=1,\cdots,K\!+\!1 $.

\subsubsection{Determine the AoA pairs corresponding to the links from  users to  semi-passive sub-IRSs}
Denote the locations of the BS and the $i$-th sub-IRS by $\mathbf{q}_{\mathrm{BS}}=\left( x_{\mathrm{BS}},y_{\mathrm{BS}},z_{\mathrm{BS}} \right) $ and $\mathbf{q}_i=\left( x_i,y_i,z_i \right) $, respectively. Due to the fixed locations of the BS and the sub-IRS, we assume that their locations are perfectly known. Therefore, the effective AoAs from the passive sub-IRS to the $i$-th ($i$ = 2, 3) sub-IRS can be calculated as
\begin{align}
&u_{\mathrm{I}2\mathrm{I},i}^{\mathrm{A}}=\frac{y_i-y_1}{\left\| \mathbf{q}_i-\mathbf{q}_1 \right\|},\,\,
v_{\mathrm{I}2\mathrm{I},i}^{\mathrm{A}}=\frac{z_i-z_1}{\left\| \mathbf{q}_i-\mathbf{q}_1 \right\|}.
\end{align}
After excluding the AoA pair corresponding to $\left( u_{\mathrm{I}2\mathrm{I},i}^{\mathrm{A}},v_{\mathrm{I}2\mathrm{I},i}^{\mathrm{A}} \right)$ from $\{ ( \hat{u}_{il}^{\left( n \right)},\hat{v}_{il}^{\left( n \right)} ) ,l=1,\cdots,K\!+\!1 \} $, the $i$-th sub-IRS obtains $K$ pairs of effective AoAs corresponding to $K$ users,   the set of which is denoted by  $\mathcal{A} _i\triangleq \{ ( \hat{u}_{il}^{(n)},\hat{v}_{il}^{(n)} ) |\,l=1,\cdots ,K \}, i=2,3$.
\subsection{Determine users' locations}
For the effective AoA pair  $( \hat{u}_{2l}^{(n)}, \hat{v}_{2l}^{(n)})$ in $\mathcal{A}_2$, we should find the effective AoA pair in $\mathcal{A}_3$, so that the two effective AoA pairs are corresponding to the same user. As such, the user location can be determined by these two AoA pairs. In the following, we first estimate the path losses  corresponding to  the $2K$ AoA pairs that belong  to $\mathcal{A}_2$ or $\mathcal{A}_3$. Then, $K^2$ possible users' locations and their corresponding path losses are computed.
Based on the estimated $2K$ path losses and the calculated  $K^2$ possible path losses, an AoA matching algorithm is proposed to determine users' locations.

\subsubsection{Estimate the path loss corresponding to each pair of effective AoAs}
According to (\ref{SMB2}) and (\ref{SMB3}), the received signal at the $i$-th sub-IRS during the $n$-th time block of the ISAC period can be rewritten as (\ref{LS10}) at the top of this page,
\begin{figure*}[t]
\begin{equation}\label{LS10}
 \begin{gathered}
  \begin{aligned}
   \mathbf{x}_{i}^{\left( n \right)}\left( t \right) &=\sqrt{\rho}\,\alpha _{\mathrm{I}2\mathrm{I},i}\mathbf{b}_i\left( u_{\mathrm{I}2\mathrm{I},i}^{\mathrm{A}},v_{\mathrm{I}2\mathrm{I},i}^{\mathrm{A}} \right) \mathbf{b}_{1}^{H}\left( u_{\mathrm{I}2\mathrm{I},i}^{\mathrm{D}},v_{\mathrm{I}2\mathrm{I},i}^{\mathrm{D}} \right) \,\!\mathbf{\Theta }_{1}^{\left( n \right)}\sum_{k=1}^K{\alpha _{\mathrm{U}2\mathrm{I},1,k}\mathbf{b}_1\left( u_{\mathrm{U}2\mathrm{I},1,k}^{\mathrm{A}},v_{\mathrm{U}2\mathrm{I},1,k}^{\mathrm{A}} \right) s_k\left( t \right)}\\
&+\sqrt{\rho}\sum_{k=1}^K{\alpha _{\mathrm{U}2\mathrm{I},i,k}\mathbf{b}_i\left( u_{\mathrm{U}2\mathrm{I},i,k}^{\mathrm{A}},v_{\mathrm{U}2\mathrm{I},i,k}^{\mathrm{A}} \right) s_k\left( t \right)}+\!\mathbf{n}_i\left( t \right) ,\,i\!=\!2,3,\,t\!\in \!\mathcal{N} _n,\,n\!=\!1,2,
 \end{aligned}
 \end{gathered}\
\end{equation}
\end{figure*}
where the first term denotes the interference from the passive sub-IRS, the second term denotes the signals from $K$ users to the $i$-th sub-IRS, and the third term is the AWGN at the $i$-th sub-IRS. 

Define $P_i\triangleq \alpha _{\mathrm{I}2\mathrm{I},i}\mathbf{b}_{1}^{H}( u_{\mathrm{I}2\mathrm{I},i}^{\mathrm{D}},v_{\mathrm{I}2\mathrm{I},i}^{\mathrm{D}} ) \,\!\mathbf{\Theta }_{1}^{\left( n \right)}\sum_{k=1}^K{\alpha _{\mathrm{U}2\mathrm{I},1,k}}
$ $\mathbf{b}_1( u_{\mathrm{U}2\mathrm{I},1,k}^{\mathrm{A}},v_{\mathrm{U}2\mathrm{I},1,k}^{\mathrm{A}} ) s_k\left( t \right) $, and (\ref{LS10}) can be compactly written as
\begin{align} \label{LS1}
\mathbf{x}_{i}^{\left( n \right)}\left( t \right) \!\!=\sqrt{\rho}\mathbf{B}_i\boldsymbol{\beta }_i\mathbf{s}+\!\mathbf{n}_i\left( t \right) ,\,i\!=\!2,3,\,t\!\in \!\mathcal{N} _n,\,n\!=\!1,2,
\end{align}
where
\begin{align}
&\mathbf{B}_i\triangleq \left[ \mathbf{b}_i\left( u_{\mathrm{U}2\mathrm{I},i,1}^{\mathrm{A}},v_{\mathrm{U}2\mathrm{I},i,1}^{\mathrm{A}} \right) ,\cdots ,\mathbf{b}_i\left( u_{\mathrm{U}2\mathrm{I},i,K}^{\mathrm{A}},v_{\mathrm{U}2\mathrm{I},i,K}^{\mathrm{A}} \right) \right.,\!\notag\\
& \qquad\qquad\qquad \left. \mathbf{b}_i\left( u_{\mathrm{I}2\mathrm{I},i}^{\mathrm{A}},v_{\mathrm{I}2\mathrm{I},i}^{\mathrm{A}} \right) \right] \in \mathbb{C} ^{M_i\times \left( K+1 \right)},\\
&\boldsymbol{\beta }_i\!\triangleq\! \mathrm{diag}\left( \alpha _{\mathrm{U}2\mathrm{I},i,1},\cdots,\alpha _{\mathrm{U}2\mathrm{I},i,K},P_i \right) \!\in\! \mathbb{C} ^{\left( K+1 \right) \times \left( K+1 \right)},\\
&\mathbf{s}\triangleq \left[ s_1\left( t \right) ,\cdots,s_K\left( t \right) ,1 \right] ^T\in \mathbb{C} ^{\left( K+1 \right) \times 1}.
\end{align}
\par Since $\left| s_k\left( t \right) \right|=1$ and $\mathbb{E} \{ s_i\left( t \right) s_j\left( t \right) \} =0, i\ne j$,  the auto-correlation matrix of $\mathbf{x}_{i}^{\left( n \right)}\!\left( t \right)$ can be expressed as
\begin{align} \label{LS2}
\mathbf{R}_{i}^{\left( n \right)}\!\triangleq \!\mathbb{E} \left\{ \mathbf{x}_{i}^{\left( n \right)}\left( t \right) \!\!\left[ \mathbf{x}_{i}^{\left( n \right)}\left( t \right) \! \right] ^H \right\} \!=\!\rho \mathbf{B}_i\mathbf{R}_{\boldsymbol{\beta }_i}\mathbf{B}_{i}^{H}\!+\!\sigma _{0}^{2}\mathbf{I},
\end{align}
where $\mathbf{R}_{\boldsymbol{\beta }_i}\!\triangleq\!\boldsymbol{\beta }_i\boldsymbol{\beta }_{i}^{H}\!=\!\mathrm{diag}( \left| \alpha _{\mathrm{U}2\mathrm{I},i,1} \right|^2\!,\cdots \!,\left| \alpha _{\mathrm{U}2\mathrm{I},i,K} \right|^2\!,| P_i |^2 ) $. 

Noticing that the path losses are determined by the diagonal elements of $\mathbf{R}_{\boldsymbol{\beta }_i}$,  we use (\ref{LS2}) and estimate $\mathbf{R}_{\boldsymbol{\beta }_i}$ in the $n$-th time block as (\ref{Rbeta}) at the top of this page,
\begin{figure*}[htbp]
\begin{equation}\label{Rbeta}
 \begin{gathered}
  \begin{aligned}
   \mathbf{\hat{R}}_{\boldsymbol{\beta }_i}^{\left( n \right)}=\frac{1}{\rho}\left( \left[ \mathbf{\hat{B}}_{i}^{\left( n \right)} \right] ^H\mathbf{\hat{B}}_{i}^{\left( n \right)} \right) ^{-1}\left[ \mathbf{\hat{B}}_{i}^{\left( n \right)} \right] ^H\left( \mathbf{\hat{R}}_{i}^{\left( n \right)}\!-\!\sigma _{0}^{2}\mathbf{I} \right) \mathbf{\hat{B}}_{i}^{\left( n \right)}\left( \left[ \mathbf{\hat{B}}_{i}^{\left( n \right)} \right] ^H\mathbf{\hat{B}}_{i}^{\left( n \right)} \right) ^{-1}, 
 \end{aligned}
 \end{gathered}\
\end{equation}
\hrulefill
\end{figure*}
where
\begin{align} 
&\mathbf{\hat{R}}_{i}^{\left( n \right)}=\frac{1}{\tau _n}\sum_{t\in \mathcal{N} _n}^{}{{\bf x}_{i}^{\left( n \right)}}(t)*\left[ {\bf x}_{i}^{\left( n \right)}(t) \right] ^H,\\
&\mathbf{\hat{B}}_{i}^{\left( n \right)}\!=\! \left[ \mathbf{b}_i\!\left( \hat{u}_{i1}^{(n)},\hat{v}_{i1}^{(n)} \right)\!,\cdots\!,\mathbf{b}_i\!\left( \hat{u}_{iK}^{(n)},\hat{v}_{iK}^{(n)} \right)\!,\mathbf{b}_i\!\left( u_{\mathrm{I}2\mathrm{I},i}^{\mathrm{A}},v_{\mathrm{I}2\mathrm{I},i}^{\mathrm{A}} \right) \right]\!\!.
\end{align}

Finally, the path loss $| \hat{\alpha}_{il}^{\left( n \right)} |$ corresponding to $( u_{il}^{(n)},v_{il}^{(n)} )$ is determined by the $l$-th diagonal element of $\mathbf{\hat{R}}_{\boldsymbol{\beta }_i}^{\left( n \right)}$.
\subsubsection{Estimate all possible users' locations and their corresponding path losses}
By combining the estimated AoA pairs $(\hat{u}_{2l}^{(n)},\hat{v}_{2l}^{(n)} ) \in \mathcal{A}_2$ and $(\hat{u}_{2s}^{(n)},\hat{v}_{2s}^{(n)} ) \in \mathcal{A}_3$, we calculate the possible user location, which satisfies the following equations:
\begin{align} 
&\hat{u}_{2l}^{(n)}=\frac{y_2-\hat{y}_{\mathrm{U},ls}^{\left( n \right)}}{\hat{d}_{\mathrm{U}2\mathrm{I},2,ls}^{\left( n \right)}},\,\,
\hat{v}_{2l}^{(n)}=\frac{z_2-\hat{z}_{\mathrm{U},ls}^{\left( n \right)}}{\hat{d}_{\mathrm{U}2\mathrm{I},2,ls}^{\left( n \right)}},
\\
&\hat{u}_{3s}^{(n)}=\frac{y_3-\hat{y}_{\mathrm{U},ls}^{\left( n \right)}}{\hat{d}_{\mathrm{U}2\mathrm{I},3,ls}^{\left( n \right)}},\,\,
\hat{v}_{3s}^{(n)}=\frac{z_3-\hat{z}_{\mathrm{U},ls}^{\left( n \right)}}{\hat{d}_{\mathrm{U}2\mathrm{I},3,ls}^{\left( n \right)}}.
\end{align}
Let $\mathbf{\hat{q}}_{\mathrm{U},ls}^{\left( n \right)}\!\!\triangleq\!\! [ \hat{x}_{\mathrm{U},ls}^{\left( n \right)},\hat{y}_{\mathrm{U},ls}^{\left( n \right)},\hat{z}_{\mathrm{U},ls}^{\left( n \right)} ] ^T$ and $\hat{d}_{\mathrm{U}2\mathrm{I},i,ls}^{\left( n \right)}\!\!\triangleq\!\! \| \mathbf{\hat{q}}_{\mathrm{U},ls}^{\left( n \right)}-\mathbf{q}_i \|$.
The above equations can be compactly expressed as
\begin{align}
\mathbf{\hat{A}}_{ls}^{\left( n \right)}\mathbf{\hat{z}}_{ls}^{\left( n \right)}=\mathbf{p},
\end{align}
where
\begin{align}
&\mathbf{\hat{A}}_{ls}^{\left( n \right)}\triangleq \left( \begin{matrix}
	1&		0&		\hat{u}_{2l}^{(n)}&		0\\
	0&		1&		\hat{v}_{2l}^{(n)}&		0\\
	1&		0&		0&		\hat{u}_{3s}^{(n)}\\
	0&		1&		0&		\hat{v}_{3s}^{(n)}\\
\end{matrix} \right),\\
&\mathbf{\hat{z}}_{ls}^{\left( n \right)}\triangleq \left[ \hat{y}_{\mathrm{U},ls}^{\left( n \right)},\hat{z}_{\mathrm{U},ls}^{\left( n \right)},\hat{d}_{\mathrm{U}2\mathrm{I},2,ls}^{\left( n \right)},\hat{d}_{\mathrm{U}2\mathrm{I},3,ls}^{\left( n \right)} \right] ^T,\\
&\mathbf{p}\triangleq \left[ y_{2,}z_{2,},y_{3,},z_3 \right] ^T.
\end{align}

By solving the above matrix equation, we obtain
\begin{align}
&\hat{d}_{\mathrm{U}2\mathrm{I},2,ls}^{\left( n \right)}=\frac{\hat{u}_{3s}^{(n)}\left( z_2-z_3 \right) -\hat{v}_{3s}^{(n)}\left( y_2-y_3 \right)}{\hat{u}_{3s}^{(n)}\hat{v}_{2l}^{(n)}-\hat{u}_{2l}^{(n)}\hat{v}_{3s}^{(n)}},\\
&\hat{d}_{\mathrm{U}2\mathrm{I},3,ls}^{\left( n \right)}=\frac{\hat{u}_{2l}^{(n)}\left( z_3-z_2 \right) -\hat{v}_{2l}^{(n)}\left( y_3-y_2 \right)}{\hat{u}_{2l}^{(n)}\hat{v}_{3s}^{(n)}-\hat{u}_{3s}^{(n)}\hat{v}_{2l}^{(n)}},\\
&\hat{y}_{\mathrm{U},ls}^{\left( n \right)}=y_2-\hat{u}_{2l}^{(n)}\hat{d}_{\mathrm{U}2\mathrm{I},2,l}^{\left( n \right)},\label{LS3}\\
&\hat{z}_{\mathrm{U},ls}^{\left( n \right)}=z_2-\hat{v}_{2l}^{(n)}\hat{d}_{\mathrm{U}2\mathrm{I},2,l}^{\left( n \right)}.\label{LS4}
\end{align}

Next, we calculate $\hat{x}_{\mathrm{U},ls}^{\left( n \right)}$ according to the following equation
\begin{align} 
&\left( \hat{x}_{\mathrm{U},ls}^{\left( n \right)}-x_i \right) ^2+\left( \hat{y}_{\mathrm{U},ls}^{\left( n \right)}-y_i \right) ^2+\left( \hat{z}_{\mathrm{U},ls}^{\left( n \right)}-z_i \right) ^2=\notag
\\
&\qquad\qquad\qquad\left( \hat{d}_{\mathrm{U}2\mathrm{I},i,ls}^{\left( n \right)} \right) ^2,i=2,3,
\end{align}
and obtain
\begin{align}\label{LS5}
\hat{x}_{\mathrm{U},ls}^{\left( n \right)}=\underset{\omega _2}{\mathrm{arg}}\underset{\,\,  \,\,\omega _2\in \left\{ x_2\pm d_{x,2,ls} \right\} ,\omega _3\in \left\{ x_3\pm d_{x,3,ls} \right\} \,\, }{\min}\left| \omega _2\!-\!\omega _3 \right|,
\end{align}
where $d_{x,i,ls}\triangleq \sqrt{( \hat{d}_{\mathrm{U}2\mathrm{I},i,ls}^{\left( n \right)} ) ^2\!-\!( \hat{y}_{\mathrm{U},ls}^{\left( n \right)}\!-\!y_i ) ^2\!-\!( \hat{z}_{\mathrm{U},ls}^{\left( n \right)}\!-\!z_i ) ^2}$.
\par Finally, by combining (\ref{LS3}), (\ref{LS4}) and (\ref{LS5}), we estimate $\mathbf{\hat{q}}_{\mathrm{U},ls}^{\left( n \right)}$ and calculate the corresponding distance-dependent path loss according to the log-distance path loss model \cite{jung2011wi} as
\begin{align} \label{LS6}
\left| \check{\alpha}_{i,ls}^{\left( n \right)} \right|=10^{-\frac{1}{20}\left( PL\left( d_0 \right) +10\epsilon _{\mathrm{U}2\mathrm{I}}\log \left( \frac{\left\| \mathbf{\hat{q}}_{\mathrm{U},ls}^{\left( n \right)}-\mathbf{q}_i \right\|}{d_0} \right) \right)},
\end{align}
where $d_{0}$ is the reference distance, and $\epsilon _{\mathrm{U}2\mathrm{I}}$ is the path loss exponent from users to the IRS.
\subsubsection{Simultaneous multi-user location sensing through AoA matching}
In order to match the effective AoA pairs in $\mathcal{A}_2$ with those in $\mathcal{A}_3$ so that they are corresponding to the same user, we propose an AoA matching algorithm.
\par For the effective AoA pair $( \hat{u}_{2l}^{(n)}, \hat{v}_{2l}^{(n)}) \in \mathcal{A}_2$, we denote its corresponding effective AoA pair in $\mathcal{A}_3$  by $( \hat{u}_{3s_l}^{(n)},\hat{v}_{3s_l}^{(n)} ) $. Then, we initialize $\mathcal{S} =\{ 1,\cdots ,K \}$ and estimate $s_l$ in the $l$-th iteration as
\begin{align}\label{SL}
\hat{s}_l=\mathrm{arg}\underset{s\in \mathcal{S}}{\min}\left| \left[ \left| \check{\alpha}_{2,ls}^{\left( n \right)} \right|,\left| \check{\alpha}_{3,ls}^{\left( n \right)} \right| \right] ^T-\left[ \left| \hat{\alpha}_{2l}^{\left( n \right)} \right|,\left| \hat{\alpha}_{3s}^{\left( n \right)} \right| \right] ^T \right|,
\end{align}
based on which, we obtain an estimated user location $\mathbf{\hat{q}}_{\mathrm{U},l\hat{s}_l}^{\left( n \right)}$, and remove $\hat{s}_l$ from $\mathcal{S}$. After $K$ iterations, we obtain the locations of $K$ users, i.e., $\mathcal{K} ^{\left( n \right)}=\{ \mathbf{\hat{q}}_{\mathrm{U},k}^{\left( n \right)},k=1,\cdots ,K \} $. The detailed process of the AoA matching algorithm is provided in Algorithm 1.
\par Finally, we summarize in Algorithm 2 the main procedures of the proposed multi-user location sensing algorithm.
\begin{remark}
	In addition to supporting localization services, the sensed users' locations can be used for improving communication performance through for example beamforming design.
\end{remark}
\begin{algorithm}[h]
\caption{AoA Matching Algorithm} 
\hspace*{0.02in} {\bf Input:}
$\left| \hat{\alpha}_{il}^{\left( n \right)} \right|$, $\left| \check{\alpha}_{i,ls}^{\left( n \right)} \right|$, $\mathbf{\hat{q}}_{\mathrm{U},ls}^{\left( n \right)}$.
\begin{algorithmic}[1]
\State Initialize $\mathcal{K} ^{\left( n \right)}=\oslash$, $\mathcal{S} =\left\{ 1,\cdots ,K \right\}$ .
\For{$l=1,\cdots,K$}
    \State Calculate $\hat{s}_l$ according to (\ref{SL}).
    \State Add $\mathbf{\hat{q}}_{\mathrm{U},l\hat{s}_l}^{\left( n \right)}$ into $\mathcal{K} ^{\left( n \right)}$.
    \State Remove $\hat{s}_l$ from $\mathcal{S}$.
\EndFor
\end{algorithmic}
\hspace*{0.02in} {\bf Output:} 
$\mathcal{K} ^{\left( n \right)}$.
\end{algorithm}

\begin{algorithm}[h]
\caption{Multi-User Location Sensing Algorithm}
\hspace*{0.02in} {\bf Input:}  
$\mathbf{x}_{i}^{\left( n \right)}\left( t \right)$, $\mathbf{q}_i$.
\begin{algorithmic}[1]
\State Estimate the effective AoA pairs corresponding to the links from the $K$ users to the two semi-passive sub-IRSs according to Section III.A, which yields  $\mathcal{A}_i,i=2,3$.
\State Estimate the path losses corresponding to the effective AoA pairs in $\mathcal{A}_i, i=2,3$ according to Section III.B.1), which yields  $| \hat{\alpha}_{il}^{(n)}  |,l=1,...,K, i=2,3$. 
\State Estimate all possible users' locations and their corresponding path losses according to Section III.B.2), which yields $\hat{\bf q}_{\text{U},ls}^{(n)}$ and  $| \check{\alpha}_{i,ls}^{(n)}  |, l,s=1,...,K, i=2,3$. 
\State Determine users' locations  $\mathcal{K}^{(n)}$ based on the Algorithm 1.
\end{algorithmic}
\hspace*{0.02in} {\bf Output:}
$\mathcal{K} ^{\left( n \right)}$.
\end{algorithm}
\vspace{-0.2cm}

\section{ Sensing-based Joint Active and Passive Beamforming}\label{section4}
In this section, we propose two sensing-based  beamforming algorithms to maximize the sum rate for the ISAC and PC periods, respectively, by capitalizing on the users' locations sensed  in the ISAC period.
\vspace{-0.2cm}
\subsection{ISAC Period}
In the ISAC period, we first design the BS combining vectors by adopting the  maximum ratio combining (MRC) technique, and then propose a cross-entropy (CE) algorithm to optimize the phase shift matrix of the passive sub-IRS.
\par During the $n$-th time block of the ISAC period, the sum rate of $K$ users is given by
\begin{align}\label{BFd}
&R_{\mathrm{sum}}\left( \mathbf{W}^{\left( n \right)},\mathbf{\Theta }_{1}^{\left( n \right)} \right)=\\
&\sum_{k=1}^K{\log _2\!\!\left( 1\!+\!\frac{\rho \left| \left[ \mathbf{w}_{k}^{\left( n \right)} \right] ^H\mathbf{H}_{\mathrm{I}2\mathrm{B},1}\mathbf{\Theta }_{1}^{\left( n \right)}\mathbf{h}_{\mathrm{U}2\mathrm{I},1,k}^{\left( n \right)} \right|^2}{\rho \!\sum\nolimits_{j\ne k}^K{\left| \left[ \mathbf{w}_{k}^{\left( n \right)} \right] ^H\mathbf{H}_{\mathrm{I}2\mathrm{B},1}\mathbf{\Theta }_{1}^{\left( n \right)}\mathbf{h}_{\mathrm{U}2\mathrm{I},1,j}^{\left( n \right)} \right|^2}\!\!+\!\sigma _{0}^{2}} \right)},\notag
\end{align}
where the combining matrix $\mathbf{W}^{\left( n \right)}$ is defined as $\mathbf{W}^{\left( n \right)}\triangleq [ \mathbf{w}_{1}^{\left( n \right)},\cdots,\mathbf{w}_{K}^{\left( n \right)} ] $. As such, the sum rate maximization problem is formulated as
\begin{subequations} 
\begin{align}\label{p1}
\text {(P1)}:&\max_{\mathbf{W}^{\left( n \right)},\mathbf{\Theta }_{1}^{\left( n \right)}}\ R_{\mathrm{sum}}\left( \mathbf{W}^{\left( n \right)},\mathbf{\Theta }_{1}^{\left( n \right)} \right) ,   \\
&\,\,\,\,\,\,\text{s.t.}\,\,\,\,\,\,\,\,\left\| \mathbf{w}_{k}^{\left( n \right)} \right\| =1,\\
&\,\,\,\,\,\,\,\,\,\,\,\,\,\,\,\,\,\,\,\,\,\,\vartheta _{1,m}^{\left( n \right)}\in \mathcal{F} ,\forall m=1,\cdots,M_{1}.
\end{align}
\end{subequations}
\subsubsection{BS combining vectors optimization}
With the given phase shift matrix $\mathbf{\Theta }_{1}^{\left( n \right)}$, the problem (P1) can be reduced to the optimization of $\mathbf{W}^{\left( n \right)}$. By adopting the MRC technique, we obtain the combining vector for the $k$-th user as
\begin{align}
\mathbf{w}_{k}^{\left( n \right)}=\frac{\mathbf{H}_{\mathrm{I}2\mathrm{B},1}\mathbf{\Theta }_{1}^{\left( n \right)}\mathbf{h}_{\mathrm{U}2\mathrm{I},1,k}^{\left( n \right)}}{\left\| \mathbf{H}_{\mathrm{I}2\mathrm{B},1}\mathbf{\Theta }_{1}^{\left( n \right)}\mathbf{h}_{\mathrm{U}2\mathrm{I},1,k}^{\left( n \right)} \right\|}.
\end{align}

Since the locations of the BS and all the sub-IRSs are fixed, we consider that ${\bf H} _{\mathrm{I}2\mathrm{B},i}$ is perfectly known \cite{5535154}. Although ${\bf h}_{\text{U2I},i,k}^{(n)}$ is unavailable due to the phase ambiguity of its complex channel gain $\alpha_{\text{U2I},i,k}$, this phase ambiguity would not effect the calculation of $R_{\mathrm{sum}}$. Therefore, we define $\mathbf{h}_{\mathrm{abs} ,i,k}^{\left( n \right)}\triangleq  | \alpha _{\mathrm{U}2\mathrm{I},i,k}^{ ( n  )} |\mathbf{b}_i( u_{\mathrm{U}2\mathrm{I},i,k}^{\mathrm{A},\left( n \right)},v_{\mathrm{U}2\mathrm{I},i,k}^{\mathrm{A},\left( n \right)} )$,
and design $\mathbf{w}_{k}^{\left( n \right)}$ as (\ref{w1}) at the top of the next page.
\begin{figure*}[htbp]
\begin{equation}\label{w1}
 \begin{gathered}
  \begin{aligned}
   \mathbf{w}_{k}^{\left( n \right)}=\frac{\mathbf{H}_{\mathrm{I}2\mathrm{B},1}\mathbf{\Theta }_{1}^{\left( n \right)}\mathbf{h}_{\mathrm{abs} ,1,k}^{\left( n \right)}}{\left\| \mathbf{H}_{\mathrm{I}2\mathrm{B},1}\mathbf{\Theta }_{1}^{\left( n \right)}\mathbf{h}_{\mathrm{abs} ,1,k}^{\left( n \right)} \right\|}=\frac{\alpha _{\mathrm{I}2\mathrm{B},1}\mathbf{a}\left( u_{\mathrm{I}2\mathrm{B},1}^{\mathrm{A}} \right) \mathbf{b}_{1}^{H}\left( u_{\mathrm{I}2\mathrm{B},1}^{\mathrm{D}},v_{\mathrm{I}2\mathrm{B},1}^{\mathrm{D}} \right) \mathbf{\Theta }_{1}^{\left( n \right)}\left| \alpha _{\mathrm{U}2\mathrm{I},1,k}^{\left( n \right)} \right|\mathbf{b}_1\left( u_{\mathrm{U}2\mathrm{I},1,k}^{\mathrm{A},\left( n \right)},v_{\mathrm{U}2\mathrm{I},1,k}^{\mathrm{A},\left( n \right)} \right)}{\left\| \mathbf{H}_{\mathrm{I}2\mathrm{B},1}\mathbf{\Theta }_{1}^{\left( n \right)}\mathbf{h}_{\mathrm{abs} ,1,k}^{\left( n \right)} \right\|}.
 \end{aligned}
 \end{gathered}\
\end{equation}
\begin{equation}\label{w2}
 \begin{gathered}
  \begin{aligned}
  \frac{\alpha _{\mathrm{I}2\mathrm{B},1}\mathbf{b}_{1}^{H}\left( u_{\mathrm{I}2\mathrm{B},1}^{\mathrm{D}},v_{\mathrm{I}2\mathrm{B},1}^{\mathrm{D}} \right) \mathbf{\Theta }_{1}^{\left( n \right)}\left| \alpha _{\mathrm{U}2\mathrm{I},1,k}^{\left( n \right)} \right|\mathbf{b}_1\left( u_{\mathrm{U}2\mathrm{I},1,k}^{\mathrm{A},\left( n \right)},v_{\mathrm{U}2\mathrm{I},1,k}^{\mathrm{A},\left( n \right)} \right)}{\left\| \mathbf{H}_{\mathrm{I}2\mathrm{B},1}\mathbf{\Theta }_{1}^{\left( n \right)}\mathbf{h}_{\mathrm{abs} ,1,k}^{\left( n \right)} \right\|}=\frac{1}{N}.
  \end{aligned}
 \end{gathered}\
\end{equation}
\hrulefill
\end{figure*}
Due to $\| \mathbf{w}_{k}^{\left( n \right)} \| =1$, (\ref{w2}) at the top of the next page can be easily verified.
Substituting (\ref{w2}) into (\ref{w1}) yields $\mathbf{w}_{k}^{(n)}\!=\!\frac{1}{N}\mathbf{a}(u_{\mathrm{I}2\mathrm{B},1}^{\mathrm{A}} ) $. Therefore, for any $\mathbf{\Theta }_{1}^{\left( n \right)}$, we obtain $\mathbf{W}^{\left( n \right)}$ as
\begin{align} \label{WISAC}
\mathbf{W}^{\left( n \right)}=\frac{1}{N}\left[ \mathbf{a}\left( u_{\mathrm{I}2\mathrm{B},1}^{\mathrm{A}} \right) ,\cdots ,\mathbf{a}\left( u_{\mathrm{I}2\mathrm{B},1}^{\mathrm{A}} \right) \right] \in \mathbb{C} ^{N\times K}.
\end{align}
\subsubsection{IRS phase shift matrix optimization}
With the given combining matrix $\mathbf{W}^{\left( n \right)}$, the problem (P1) can be simplified to the optimization of $\mathbf{\Theta }_{1}^{\left( n \right)}$. Since the user location information is unavailable in the first time block, $\mathbf{\Theta }_{1}^{\left( 1 \right)}$ is randomly generated. Therefore, in the following, we only focus on the design of $\mathbf{\Theta }_{1}^{\left( 2 \right)}$ in the second time block by invoking the users' locations sensed in the first time block. 
\par First, we set the probability matrix corresponding to $\mathbf{\Theta }_{1}^{\left( 2 \right)}$ as $\mathbf{p}_{1}^{\left( 2 \right)}=[ \mathbf{p}_{1,1}^{\left( 2 \right)},\cdots,\mathbf{p}_{1,m}^{\left( 2 \right)},\cdots,\mathbf{p}_{1,M_1}^{\left( 2 \right)} ] \in \mathbb{C} ^{2^b\times M_1}$, where $\mathbf{p}_{1,m}^{\left( 2 \right)}=[ p_{1,m,1}^{\left( 2 \right)},\cdots,p_{1,m,2^b}^{\left( 2 \right)} ] ^T$
denotes the probability parameter for $\vartheta _{1,m}^{\left( 2 \right)}$, with its entry $p_{1,m,l}^{\left( 2 \right)}$ satisfying the probability constraints $0\leqslant p_{1,m,l}^{\left( 2 \right)}\leqslant 1$ and $\sum_{l=1}^{2^b}{}p_{1,m,l}^{\left( 2 \right)}=1$. 
Subsequently, we consider that $\vartheta _{1,m}^{\left( 2 \right)}$ takes a value from $\mathcal{F} $ with an equal probability at first, and initialize the  probability matrix as $\mathbf{p}_{1}^{\left( 2 \right) ,0}\!=\!\frac{1}{2^b}\times \mathbf{1}_{2^b\times M_1}$. Then, in the $i$-th iteration, we randomly generate $S_{\mathrm{ISAC}}$ candidates $\{ \mathbf{\Theta }_{1}^{\left( 2 \right) ,s} \} _{s=1}^{S_{\mathrm{ISAC}}}$ according to the probability distribution function given by
\begin{align} 
\Xi \left( \mathbf{\Theta }_{1}^{\left( 2 \right)};\mathbf{p}_{1}^{\left( 2 \right) ,i} \right)& =\prod_{m=1}^{M_1}{\left( \left( \prod_{l=1}^{2^b-1}{\left( p_{1,m,l}^{\left( 2 \right) ,i} \right) ^{\varGamma \left( \vartheta _{1,m}^{\left( 2 \right)},\mathcal{F} \left( l \right) \right)}} \right) \right.}\notag\\
&\!\!\times \!\left. \left( 1\!-\!\prod_{l=1}^{2^b-1}{\left( p_{1,m,l}^{\left( 2 \right) ,i} \right) ^{\varGamma \left( \vartheta _{1,m}^{\left( 2 \right)},\mathcal{F} \left( l \right) \right)}} \right) \right)\!,
\end{align}
where $\mathcal{F} \left( l \right) $ denotes the $l$-th entry of $\mathcal{F}$, and $\varGamma ( \vartheta _{1,m}^{\left( 2 \right)},\mathcal{F} \left( l \right) ) $ is a judge function  given by
\begin{align} 
\varGamma \left( \vartheta _{1,m}^{\left( 2 \right)},\mathcal{F} \left( l \right) \right) =\begin{cases}
	1, \vartheta _{1,m}^{\left( 2 \right)}=\mathcal{F} \left( l \right)\\
	0, \vartheta _{1,m}^{\left( 2 \right)}\ne \mathcal{F} \left( l \right)\\
\end{cases}.
\end{align}
\par For each phase shift matrix candidate ${\bm \Theta}_{1}^{(2),s}$, we calculate the corresponding sum rate
\begin{align}\label{BF1}
&R_{\mathrm{sum}}\left( \mathbf{\Theta }_{1}^{\left( 2 \right) ,s} \right) =\\
&\sum_{k=1}^K{\!\log _2\!\left(\! 1\!+\!\frac{\rho \left| \left[ \mathbf{w}_{k}^{\left( 2 \right)} \right] ^H\!\!\mathbf{H}_{\mathrm{I}2\mathrm{B},1}\mathbf{\Theta }_{1}^{\left( 2 \right) ,s}\mathbf{\hat{h}}_{\mathrm{abs} ,1,k}^{\left( 1 \right)} \right|^2}{\rho\! \sum\nolimits_{j\ne k}^K{\left| \left[ \mathbf{w}_{k}^{\left( 2 \right)} \right] ^H\!\!\mathbf{H}_{\mathrm{I}2\mathrm{B},1}\mathbf{\Theta }_{1}^{\left( 2 \right) ,s}\mathbf{\hat{h}}_{\mathrm{abs} ,1,j}^{\left( 1 \right)} \right|^2}\!\!+\!\sigma _{0}^{2}} \right)},\notag
\end{align}
where
\begin{align}
\mathbf{\hat{h}}_{\mathrm{abs} ,1,k}^{\left( n \right)}=\left| \hat{\alpha}_{\mathrm{U}2\mathrm{I},1,k}^{\left( n \right)} \right|\mathbf{b}_1\left( \hat{u}_{\mathrm{U}2\mathrm{I},1,k}^{\mathrm{A},\left( n \right)},\hat{v}_{\mathrm{U}2\mathrm{I},1,k}^{\mathrm{A},\left( n \right)} \right), n=1,2,
\end{align}
where $| \hat{\alpha}_{\mathrm{U}2\mathrm{I},1,k}^{\left( n \right)} |\!\!=\!\!10^{-\frac{1}{20}( PL\left( d_0 \right) +10\epsilon _{\mathrm{U}2\mathrm{I}}\log ( \frac{\| \mathbf{\hat{q}}_{\mathrm{U},k}^{\left( n \right)}-\mathbf{q}_1 \|}{d_0} )\! )}$, $\hat{u}_{\mathrm{U}2\mathrm{I},1,k}^{\mathrm{A},\left( n \right)}$ $\!=\!\frac{\hat{y}_{\mathrm{U},k}^{\left( n \right)}-y_1}{\| \mathbf{\hat{q}}_{\mathrm{U},k}^{\left( n \right)}-\mathbf{q}_1 \|}$, and $\hat{v}_{\mathrm{U}2\mathrm{I},1,k}^{\mathrm{A},\left( n \right)}\!=\!\frac{\hat{z}_{\mathrm{U},k}^{\left( n \right)}-z_1}{\| \mathbf{\hat{q}}_{\mathrm{U},k}^{\left( n \right)}-\mathbf{q}_1 \|}$.
\par After that, we sort $\mathcal{R} _{\mathrm{ISAC}} \!=\!\{ R_{\mathrm{sum}}( \mathbf{\Theta }_{1}^{\left( 2 \right) ,s} ) \} _{s=1}^{S_{\mathrm{ISAC}}}$ in descending order and select the $S_{\mathrm{ISAC}}^{\mathrm{elite}}$ phase shift matrix samples corresponding to the $S_\text{ISAC}^\text{elite}$ largest sum rates in $\mathcal{R} _{\mathrm{ISAC}}$, i.e., $ {\bm \Theta}_{1}^{(2),s_q}, s_q=1,\cdots ,S_{\mathrm{ISAC}}^{\mathrm{elite}}$. Based on the selected  $S_\text{ISAC}^\text{elite}$ samples, we update the probability matrix in the $(i+1)$-th iteration as $\mathbf{p}_{1}^{\left( 2 \right) ,i+1}$ with its elements given by
\begin{align} \label{BF2}
p_{1,m,l}^{\left( 2 \right) ,i+1}=&\frac{1}{S_{\mathrm{ISAC}}^{\mathrm{elite}}}\sum_{s_q=1}^{S_{\mathrm{ISAC}}^{\mathrm{elite}}}{\varGamma \left( \vartheta _{1,m}^{\left( 2 \right) ,s_q},\mathcal{F} \left( l \right) \right)},\notag\\
&m=1,\cdots ,M_1, \,l=1,\cdots ,2^b.
\end{align}
\par Repeat the above process until the difference between the maximum and minimum of $\mathcal{R} _{\mathrm{ISAC}}$ is less than the threshold $\kappa$, which indicates that the probability matrix corresponding to $\mathbf{\Theta }_{1}^{\left( 2 \right)}$ is stable. The procedures of the beamforming algorithm in the second time block of the ISAC period is summarized in Algorithm 3.
\begin{algorithm}[h]
\caption{Sensing-Based Beamforming for the Second Time Block of the ISAC Period}
\hspace*{0.02in} {\bf Input:}  
$\mathcal{K} ^{\left( 1 \right)}$, $\mathbf{H}_{\mathrm{I}2\mathrm{B},1}$, $b$, $S_{\mathrm{ISAC}}$, $S_{\mathrm{ISAC}}^{\mathrm{elite}}$.
\begin{algorithmic}[1]
\State Initialize $\mathcal{F} =\left\{ 0,\frac{2\pi}{2^b},\cdots,\frac{2\pi}{2^b}\left( 2^b-1 \right) \right\} $, $\mathbf{p}_{1}^{\left( 2 \right) ,0}=\frac{1}{2^b}\times \mathbf{1}_{2^b\times M_1}$,
 and $i=1$.
\State Calculate $\mathbf{W}^{\left( n \right)}$ based on the MRC method.
\Repeat
    \State Randomly generate $S$ candidates $\{ \mathbf{\Theta }_{1}^{\left( 2 \right) ,s} \} _{s=1}^{S_{\mathrm{ISAC}}}$ based on $\Xi ( \mathbf{\Theta }_{1}^{\left( 2 \right)};\mathbf{p}_{1}^{\left( 2 \right) ,i} ) $.
    \State Calculate the sum rate $\{ R_{\mathrm{sum}}( \mathbf{\Theta }_{1}^{\left( 2 \right) ,s} ) \} _{s=1}^{S_{\mathrm{ISAC}}}$ by (\ref{BF1}).
    \State Sort $\mathcal{R} _{\mathrm{ISAC}} =\{ R_{\mathrm{sum}}( \mathbf{\Theta }_{1}^{\left( 2 \right) ,s} ) \} _{s=1}^{S_{\mathrm{ISAC}}}$ in descending order.
    \State Select $S_{\mathrm{ISAC}}^{\mathrm{elite}}$ phase shift matrix samples\! corresponding \!to\! the\! $S_{\mathrm{ISAC}}^{\mathrm{elite}}$ largest sum rates in $\mathcal{R} _{\mathrm{ISAC}}$.
    \State Update $\mathbf{p}_{1}^{\left( 2 \right) ,i+1}$ according to (\ref{BF2}).
    \State Let $i \rightarrow i+1 $.
\Until $\left| \max \left( \mathcal{R} _{\mathrm{ISAC}} \right) -\min \left( \mathcal{R} _{\mathrm{ISAC}} \right) \right|<\kappa $
\end{algorithmic}
\hspace*{0.02in} {\bf Output:}
$\mathbf{\Theta }_{1}^{\left( 2 \right) ,\mathrm{opt}}\!\!=\!\mathrm{diag}( \boldsymbol{\xi }_{1}^{\left( 2 \right) ,\mathrm{opt}} ) \!=\!\mathbf{\Theta }_{1}^{(2),1}$\!, $\mathbf{W}^{\left( 2 \right),\mathrm{opt}}\!=\!{\bf W}^{(n)}$.
\end{algorithm}
\vspace{-0.5cm}
\subsection{PC Period}
\par During the PC period, all three sub-IRSs operate in the reflecting mode to assist uplink transmission, and the sum rate of $K$ users can be expressed as
\vspace{-0.1cm}
\begin{align} 
&R_{\mathrm{sum}}\left( \mathbf{W}\left( t \right) ,\mathbf{\Theta }\left( t \right) \right)=\\
&\sum_{k=1}^K{\log _2\left( 1+\frac{\rho \left| \mathbf{w}_{k}^{H}\left( t \right) \mathbf{H}_{\mathrm{I}2\mathrm{B}}\mathbf{\Theta }\left( t \right) \mathbf{h}_{\mathrm{U}2\mathrm{I},k} \right|^2}{\rho \sum\nolimits_{j\ne k}^K{\left| \mathbf{w}_{k}^{H}\left( t \right) \mathbf{H}_{\mathrm{I}2\mathrm{B}}\mathbf{\Theta }\left( t \right) \mathbf{h}_{\mathrm{U}2\mathrm{I},j} \right|^2}+\sigma _{0}^{2}} \right)},\notag\\
&\qquad\qquad\qquad\qquad\qquad\quad t \in \mathcal{T}_2,\notag
\end{align}
where $\mathbf{W}\left( t \right)$ is defined as $\mathbf{W}\left( t \right) \triangleq \left[ \mathbf{w}_1\left( t \right) ,\cdots,\mathbf{w}_K\left( t \right) \right]$. As such, the sum rate maximization problem is formulated as

\begin{subequations} 
\begin{align}
\text {(P2)}:&\max_{\mathbf{W}\left( t \right)  ,\mathbf{\Theta }\left( t \right) }\ R_{\mathrm{sum}}\left( \mathbf{W}\left( t \right) ,\mathbf{\Theta }\left( t \right) \right) ,   \\
&\,\,\,\,\,\,\text{s.t.}\,\,\,\,\,\left\| \mathbf{w}_k\left( t \right) \right\| =1,\\
&\,\,\,\,\,\,\,\,\,\,\,\,\,\,\,\,\,\,
\vartheta _{i,m}\!\in\! \mathcal{F} ,\forall m\!=\!1,\cdots,M_i,\forall i\!=\!1,2,3.\end{align}
\end{subequations}
\par Note that the calculation of the sum rate involves ${\bf h}_{\text{U2I},k}$, which is unavailable due to the phase ambiguity of $\alpha _{\mathrm{U}2\mathrm{I},i,k}$. Hence, during the first $C$ time slots of the PC period ($C \ll T_2$), we set $\mathbf{W}\left( t \right) = \mathbf{W}^{\left( 2 \right) ,\mathrm{opt}}$ and $\mathbf{\Theta }\left( t \right) =\mathrm{diag}\left( \boldsymbol{\xi }\left( t \right) \right) $ in time slot $t\in \mathcal{T} _{\mathrm{c}}\triangleq \left\{ T_1+1,\cdots ,T_1+C \right\} $, where $\boldsymbol{\xi }\left( t \right) =[ [ \boldsymbol{\xi }_{1}^{\left( 2 \right) ,\mathrm{opt}} ] ^T,\left[ \boldsymbol{\xi }_2(t) \right] ^T,\left[ \boldsymbol{\xi }_3(t) \right] ^T ] ^T$, and $\boldsymbol{\xi }_i(t)$ is randomly generated phase shift beam of the $i$-th ($i=2, 3$) sub-IRS. 

\par With the received signal strength obtained in the first $C$ time slots of the PC period and the users' locations estimated in the ISAC period,  the phase ambiguity of $ \alpha_{\text{U2I},i,k}$ can be removed, which enables the sum rate calculation  in the remaining  time slots of the PC period.  As such, in the remaining  time slots of the PC period, we propose a  joint active and passive beamforming algorithm to solve the sum rate maximization problem (P2), by combining the use of zero-forcing (ZF) and CE-based techniques.



\subsubsection{Remove the phase ambiguity of $ \alpha_{\text{U2I},i,k}$}
\par We first reformulate ${\bf h}_{\text{U2I},k}$ as
\begin{align} 
&\mathbf{h}_{\mathrm{U}2\mathrm{I},k}=\left[ e^{j\psi _{1,k}}\mathbf{h}_{\mathrm{abs} ,1,k}^{T},\,\,e^{j\psi _{2,k}}\mathbf{h}_{\mathrm{abs} ,2,k}^{T},\,\,e^{j\psi _{3,k}}\mathbf{h}_{\mathrm{abs} ,3,k}^{T} \right] ^T\notag
\\
&\!\!=\!e^{j\psi _{1,k}}\!\!\left[ \mathbf{h}_{\mathrm{abs} ,1,k}^{T},\,e^{j\left( \psi _{2,k}\!-\!\psi _{1,k} \right)}\mathbf{h}_{\mathrm{abs} ,2,k}^{T},\,e^{j\left( \psi _{3,k}\!-\!\psi _{1,k} \right)}\mathbf{h}_{\mathrm{abs} ,3,k}^{T} \right] ^T\notag
\\
&\!\!=\!e^{j\psi _{1,k}}\!\!\left[ \mathbf{h}_{\mathrm{abs} ,1,k}^{T},\,e^{j\varDelta _{2,k}}\mathbf{h}_{\mathrm{abs} ,2,k}^{T},\,e^{j\varDelta _{3,k}}\mathbf{h}_{\mathrm{abs} ,3,k}^{T} \right] ^T,
\end{align}
where $\mathbf{h}_{\mathrm{abs} ,i,k}\triangleq \left| \alpha _{\mathrm{U}2\mathrm{I},i,k} \right|\mathbf{b}_i( u_{\mathrm{U}2\mathrm{I},i,k}^{\mathrm{A}},v_{\mathrm{U}2\mathrm{I},i,k}^{\mathrm{A}} ) $ and $\psi _{i,k}$ denotes the phase of $\alpha _{\mathrm{U}2\mathrm{I},i,k}$. In the following, we try to estimate $\varDelta _{i,k}$ ($i\in \left\{ 2,3 \right\} $ and $ k\in \left\{ 1,\cdots ,K \right\} $), 
by exploiting the BS received signal strength obtained during the first $C$ time slots of the PC period. Define 
\begin{align}
\mathbf{D}\triangleq \left[ \begin{array}{c}
	\varDelta _{2,1},\cdots ,\varDelta _{2,k},\cdots ,\varDelta _{2,K}\\
	\varDelta _{3,1},\cdots ,\varDelta _{3,k},\cdots ,\varDelta _{3,K}\\
\end{array} \right],
\end{align}
and assume that $\varDelta _{i,k}$ belongs to the set $\mathcal{F} _{\varDelta}=\{ 0,\frac{2\pi}{2^{b_{\varDelta}}},\cdots ,\frac{2\pi}{2^{b_{\varDelta}}}( 2^{b_{\varDelta}}-1 ) \} $, where $b_{\varDelta}$ is the bit-quantization number, and $\varDelta _{i,k}$ is approximately a continuous phase shift when $b_{\varDelta}$ is sufficiently large. Based on this assumption, $\mathbf{D}$ has $4^{Kb_{\varDelta}}$ possible values ( i.e., $\mathcal{D} \triangleq \left\{ \mathbf{D}\,|\,\varDelta _{i,k}\in \mathcal{F} _{\varDelta},i=2,3, \right.$ $\left. k=1,\cdots ,K \right\} $). For each possible value $\mathbf{\check{D}}\in \mathcal{D} $, the corresponding BS received signal strength in time slot $t\in \mathcal{T}_{\text c}$ can be calculated as
\begin{align}
\check{P}\left( \mathbf{\check{D}},\mathbf{\Theta }\left( t \right) \right) =&\sum_{k=1}^K{\left( \rho \left| \mathbf{w}_{k}^{H}\left( t \right) \mathbf{H}_{\mathrm{I}2\mathrm{B}}\mathbf{\Theta }\left( t \right) \mathbf{\check{h}}_{\varDelta ,k} \right|^2 \right)}\notag\\
&\,\,+\sigma _{0}^{2},  t \in \mathcal{T}_\text{c},
\end{align}
where $\mathbf{\check{h}}_{\varDelta ,k}\!=\![ [ \mathbf{\hat{h}}_{\mathrm{abs} ,1,k}^{\left( 2 \right)} ] ^T\!,e^{j\check{\varDelta}_{2,k}}\![ \mathbf{\hat{h}}_{\mathrm{abs} ,2,k}^{\left( 2 \right)} ] ^T\!,e^{j\check{\varDelta}_{3,k}}\![ \mathbf{\hat{h}}_{\mathrm{abs} ,3,k}^{\left( 2 \right)} ] ^T ]^T\!.$

Then, compare the possible BS received signal strength $\check{P}\left( \mathbf{\check{D}},\mathbf{\Theta }\left( t \right) \right) $ with the BS received signal strength $P\left( \mathbf{\Theta }\left( t \right) \right) $ obtained during the first $C$ time slots of the PC period, and define
\begin{align}
f\left( \mathbf{\check{D}} \right) \triangleq \sum_{t\in \mathcal{T} _{\mathrm{c}}}{\left| \check{P}\left( \mathbf{\check{D}},\mathbf{\Theta }\left( t \right) \right) -P\left( \mathbf{\Theta }\left( t \right) \right) \right|}.
\end{align}
As such, $\varDelta _{i,k}$ can be estimated as
\begin{align}\label{HU2I0}
\hat{\varDelta}_{i,k}\!=\!\left[ \underset{}{\mathrm{arg}}\,\,\underset{\mathbf{\check{D}}\in \mathcal{D}}{\min}f\left( \mathbf{\check{D}} \right) \right] _{\left( i-1 \right) k},i\!=\!2,3,\,k\!=\!1,\cdots ,K.
\end{align}
\subsubsection{Joint active and passive beamforming}
\par First, let $\mathbf{\Theta }\left( t \right) =\mathrm{diag}( e^{j\theta _1},\cdots ,e^{j\theta _m},\cdots ,e^{j\theta _M} )$ and set the corresponding probability matrix as $\mathbf{p}=[ \mathbf{p}_1,\cdots,\mathbf{p}_m,\cdots,\mathbf{p}_M ] \in \mathbb{C} ^{2^b\times M}$, where $\mathbf{p}_m=[ p_{m,1},\cdots,p_{m,2^b} ] ^T$
denotes the probability parameter for $\vartheta _m$, with its entry $p_{m,l}$ satisfying the probability constraints $0\leqslant p_{m,l}\leqslant 1$ and $\sum_{l=1}^{2^b}{}p_{m,l}=1$. Subsequently, before the first iteration, we consider that $\vartheta _m$ takes a value from $\mathcal{F} $ with an equal probability, and initialize $\mathbf{p}^0=\frac{1}{2^b}\times \mathbf{1}_{2^b\times M}$. Then, in the $i$-th iteration, we randomly generate $S_{\mathrm{PC}}$ candidates $\{ \mathbf{\Theta }^s\left( t \right) \} _{s=1}^{S_{\mathrm{PC}}}$ according to the probability distribution function given by
\begin{align}
\Xi \left( \mathbf{\Theta }\left( t \right) ;\mathbf{p}^i \right) &=\prod_{m=1}^M{\left( \left( \prod_{l=1}^{2^b-1}{\left( p_{m,l}^{i} \right) ^{\varGamma \left( \vartheta _m,\mathcal{F} \left( l \right) \right)}} \right) \right.}\notag\\
&\times \left. \left( 1-\prod_{l=1}^{2^b-1}{\left( p_{m,l}^{i} \right) ^{\varGamma \left( \vartheta _m,\mathcal{F} \left( l \right) \right)}} \right) \right).
\end{align}
\par For each $\mathbf{\Theta }^s\left( t \right) $, we define the  corresponding effective channel as $\mathbf{H}_{\mathrm{eq}}^{s}\triangleq \mathbf{H}_{\mathrm{I}2\mathrm{B}}\mathbf{\Theta }^s\left( t \right) \mathbf{\hat{H}}_{\varDelta}\in \mathbb{C} ^{N\times K}$,
where
\begin{align}
    &\mathbf{\hat{H}}_{\varDelta}=\left[ \mathbf{\hat{h}}_{\varDelta ,1},\cdots ,\mathbf{\hat{h}}_{\varDelta ,k},\cdots ,\mathbf{\hat{h}}_{\varDelta ,K} \right] ,\\
    &\mathbf{\hat{h}}_{\varDelta ,k}\!=\! \left[\! \left[ \mathbf{\hat{h}}_{\mathrm{abs} ,1,k}^{\left( 2 \right)} \right] ^T\!\!,e^{j\hat{\varDelta}_{2,k}}\!\left[ \mathbf{\hat{h}}_{\mathrm{abs} ,2,k}^{\left( 2 \right)} \right] ^T\!\!,e^{j\hat{\varDelta}_{3,k}}\!\left[ \mathbf{\hat{h}}_{\mathrm{abs} ,3,k}^{\left( 2 \right)} \right] ^T \right] ^T.
\end{align}

Then, following the ZF technique, we compute
\begin{align} 
\mathbf{\check{W}}^s\left( t \right)\! \triangleq\! \left[ \mathbf{\check{w}}_{1}^{s}\left( t \right) ,\cdots ,\mathbf{\check{w}}_{K}^{s}\left( t \right) \right]\! =\!\left[ \left[ \mathbf{H}_{\mathrm{eq}}^{s} \right] ^T \right] ^{\dagger}\in \mathbb{C} ^{N\times K},
\end{align}
based on which, the combining vector for the $k$-th user is calculated as
\begin{align}
\mathbf{w}_{k}^{s}\left( t \right) =\frac{\mathbf{\check{w}}_{k}^{s}\left( t \right)}{\left\| \mathbf{\check{w}}_{k}^{s}\left( t \right) \right\|}.  
\end{align}


\par As such, the sum rate for each phase shift matrix candidate $\mathbf{\Theta }^s\left( t \right) $ can be calculated as
\begin{align} \label{BF5}
&R_{\mathrm{sum}}\left( \mathbf{\Theta }^s\left( t \right) \right) =\\
&\sum_{k=1}^K{\log _2\left( 1\!+\!\frac{\rho \left| \left[ \mathbf{w}_{k}^{s}\left( t \right) \right] ^H\mathbf{H}_{\mathrm{I}2\mathrm{B}}\mathbf{\Theta }^s\left( t \right) \mathbf{\hat{h}}_{\varDelta ,k} \right|^2}{\rho\! \sum\nolimits_{j\ne k}^K{\left| \left[ \mathbf{w}_{k}^{s}\left( t \right) \right] ^H\mathbf{H}_{\mathrm{I}2\mathrm{B}}\mathbf{\Theta }^s\left( t \right) \mathbf{\hat{h}}_{\varDelta ,j} \right|^2}\!+\!\sigma _{0}^{2}} \right)}.\notag
\end{align}

Next, we sort $\mathcal{R} _{\mathrm{PC}}\!=\!\{ R_{\mathrm{sum}}( \mathbf{\Theta }^s\!\left( t \right) ) \} _{s=1}^{S_{\mathrm{PC}}}$ in descending order and select $S_{\mathrm{PC}}^{\mathrm{elite}}$ phase shift matrix samples corresponding to the $S_\text{PC}^\text{elite}$ largest sum rates in $\mathcal{R} _{\mathrm{PC}}$, i.e., $ \mathbf{\Theta }^{s_q}, s_q=1,..., S_\text{PC}^\text{elite}$. Based on the selected  $S_\text{PC}^\text{elite}$ samples, we  update the probability matrix in the $(i+1)$-th iteration as $\mathbf{p}^{i+1}$ with its elements given by
\begin{align}\label{BF6}
p_{m,l}^{i+1}=&\frac{1}{S_{\mathrm{PC}}^{\mathrm{elite}}}\sum_{s_q=1}^{S_{\mathrm{PC}}^{\mathrm{elite}}}{\varGamma \left( \vartheta _{m}^{s_q},\mathcal{F} \left( l \right) \right)},\notag\\
&m=1,\cdots ,M, \,l=1,\cdots ,2^b.
\end{align}
\par Repeat the above process until the difference between the maximum and minimum of $\mathcal{R} _{\mathrm{PC}}$ is less than the threshold $\kappa$. The detailed process of the beamforming algorithm in the PC period is given by Algorithm 4.

\begin{remark}
It is worth noting that the proposed two beamforming algorithms for both ISAC and PC periods are based on sensed location information instead of perfect CSI, which avoids high channel estimation overhead in the IRS-aided system.
\end{remark}
\begin{algorithm}[h]
\caption{Sensing-Based Beamforming Algorithm in the PC Period}
\hspace*{0.02in} {\bf Input:}  
$\mathcal{K} ^{\left( 2 \right)}$, $\mathbf{H}_{\mathrm{I}2\mathrm{B}}$, $b$, $S_{\mathrm{PC}}$, $S_{\mathrm{PC}}^{\mathrm{elite}}$.
\begin{algorithmic}[1]
\State Initialize $\mathcal{F} =\left\{ 0,\frac{2\pi}{2^b},\cdots,\frac{2\pi}{2^b}\left( 2^b-1 \right) \right\} $, $\mathbf{p}^0=\frac{1}{2^b}\times \mathbf{1}_{2^b\times M}$, and $i=1$.
\State Estimate $\varDelta _{i,k}$ according to (\ref{HU2I0}).
\Repeat
    \State Randomly generate $S_{\mathrm{PC}}$ candidates $\left\{ \mathbf{\Theta }^s\left( t \right) \right\} _{s=1}^{S_{\mathrm{PC}}}$ based on $\Xi ( \mathbf{\Theta }\left( t \right) ;\mathbf{p}^i )$.
    \State Calculate $\mathbf{W}^s\left( t \right) $ based on the ZF method.
    \State Calculate the sum rate $\left\{ R_{\mathrm{sum}}\left( \mathbf{\Theta }^s\left( t \right) \right) \right\} _{s=1}^{S_{\mathrm{PC}}}$ by (\ref{BF5}).
    \State Sort $\mathcal{R} _{\mathrm{PC}}=\left\{ R_{\mathrm{sum}}\left( \mathbf{\Theta }^s\left( t \right) \right) \right\} _{s=1}^{S_{\mathrm{PC}}}$ in descending order.
    \State Select $S_{\mathrm{PC}}^{\mathrm{elite}}$ phase shift matrix samples corresponding to the $S_{\mathrm{PC}}^{\mathrm{elite}}$ largest sum rate in $\mathcal{R} _{\mathrm{PC}}$.
    \State Update $\mathbf{p}^{i+1}$ based on (\ref{BF6}).
    \State Let $i \rightarrow i+1 $.
\Until $\left| \max \left( \mathcal{R} _{\mathrm{PC}} \right) -\min \left( \mathcal{R} _{\mathrm{PC}} \right) \right|<\kappa $
\end{algorithmic}
\hspace*{0.02in} {\bf Output:}
$\mathbf{\Theta }^{\mathrm{opt}}\left( t \right) =\mathbf{\Theta }^1\left( t \right) $, $\mathbf{W}^{\mathrm{opt}}\left( t \right) =\mathbf{W}^1\left( t \right) $
\end{algorithm}

\section{Simulation Results}\label{section5}
\begin{figure}[htbp]
  \centering
  \includegraphics[width=3.2in]{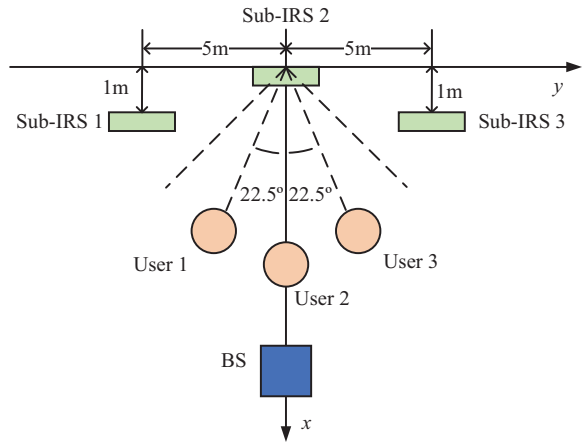}
  \caption{Simulation setup (top view).}
  \label{setup}
\end{figure}

In this section, numerical results are provided to demonstrate the effectiveness of the proposed multi-user location sensing and joint beamforming algorithms as well as to investigate the performance of the IRS-enabled multi-user ISAC system. The simulation setup is shown in Fig.~\ref{setup}, where $K$ users are distributed on a ninety-degree sector of the horizontal floor. The BS is $20$ meters (m) above the horizontal floor, and the three sub-IRSs are respectively $5$ m, $7$ m, and $9$ m above the horizontal floor.  The distances from the BS to the second sub-IRS and from the 
second sub-IRS to the users are set to be $d_{\mathrm{B}2\mathrm{I},2}=50$ m and $d_{\mathrm{I}2\mathrm{U},2}=10$ m, respectively. The path loss at the reference distance of $1$ m is set as $30$ dB. Other system
parameters are set as follows: $K=3$, $N=8$, $M_1=32\times 32$, $M_2=M_3=M_{\mathrm{semi}}=12\times 12$, $T=1200$, $T_{1}=120$, $\tau _1=20$, $\epsilon _{\mathrm{I}2\mathrm{B}}=2.3$, $\epsilon _{\mathrm{U}2\mathrm{I}}=2.2$, $\epsilon _{\mathrm{I}2\mathrm{I}}=2.1$, $b=3$, $b_{\varDelta}=4$, $S_{\mathrm{ISAC}}=1500$, $S_{\mathrm{ISAC}}^{\mathrm{elite}}=300$, $S_{\mathrm{PC}}=2000$, $S_{\mathrm{PC}}^{\mathrm{elite}}=400$, $C=4$, $\rho =20$ dBm and $\sigma _{0}^{2}=-80$ dBm (if not specified otherwise).
\subsection{Performance of Multi-User Location Sensing}
Without loss of generality, we focus on the localization during the first time block of the ISAC period. We adopt the root mean square
error (RMSE) to measure the performance of localization,
\begin{align}
\varepsilon =\mathbb{E} \left\{ \sqrt{\frac{1}{K}\sum_{k=1}^K{\left\| \mathbf{\hat{q}}_{\mathrm{U},k}-\mathbf{q}_{\mathrm{U},k} \right\| ^2}} \right\} .
\end{align}
\begin{figure}[htbp]
  \centering
  \includegraphics[width=3.2in]{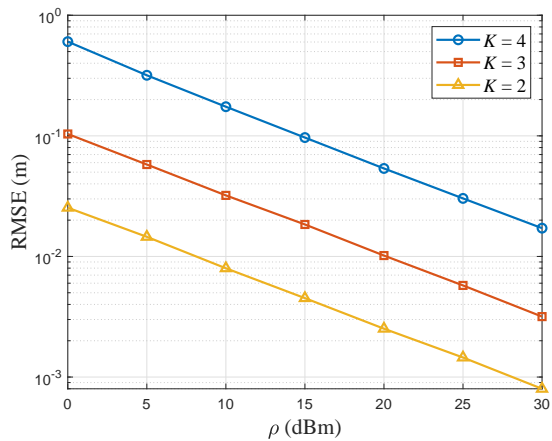}
  \caption{RMSE of the estimated users' locations versus transmit power.}
  \label{sense_rho}
\end{figure}

Fig.~\ref{sense_rho} shows the RMSE of the estimated users' locations versus the transmit power with different numbers of users. We can observe that the localization accuracy improves with the transmit power. In addition, increasing the number of users leads to a lower localization accuracy. For instance, the localization accuracy is reduced by six times as the number of users increases from 2 to 3. This is because, with a given number of semi-passive elements, the path discrimination capability of the semi-passive IRS is fixed. With more users, the semi-passive sub-IRS has to distinguish more user-IRS paths, which would reduce the accuracy of the estimated effective AoAs corresponding to each user-IRS path.

\begin{figure}[htbp]
  \centering
  \includegraphics[width=3.2in]{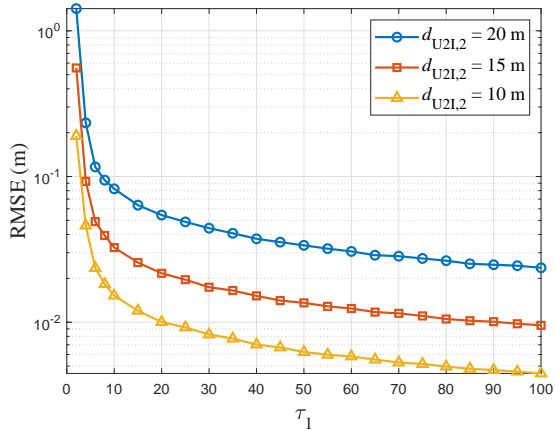}
  \caption{RMSE of the estimated users' locations versus sensing time.}
  \label{sense_tau}
\end{figure}

Fig.~\ref{sense_tau} presents the RMSE of the estimated users' locations versus the sensing time (i.e., $\tau_1$) with different user-IRS distances. For all three configurations of user-IRS distances, the RMSE of the estimated users' locations decreases as the sensing time becomes larger. This is because collecting more data can help suppress the adverse effect of noise, thereby improving the localization accuracy. Besides, the localization accuracy degrades with the increase of the user-IRS distance. This performance degradation can be compensated by increasing sensing time. For instance, when ${d}_{\mathrm{U}2\mathrm{I},2}$ increases from $10$ m to $15$ m, the RMSE remains unchanged at $10^{-2}$ by increasing $\tau_1$ from about 20 to 90.
\vspace{-0.3cm}
\begin{figure}[htbp]
  \centering
  \includegraphics[width=3.2in]{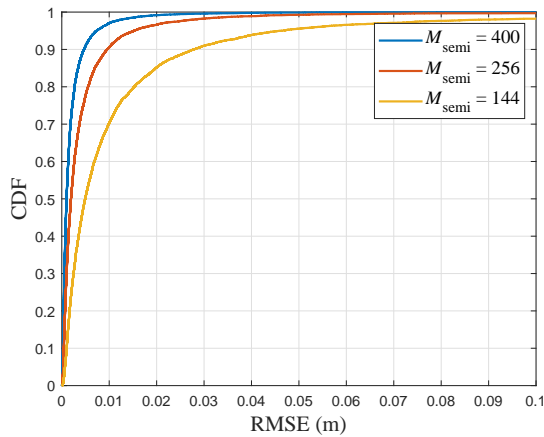}
  \caption{CDF curves of the RMSE.}
  \label{sense_CDF}
\end{figure}

Finally, in Fig.~\ref{sense_CDF}, we illustrate the cumulative distribution function of the RMSE by considering that all users are randomly distributed in a square centered at $(4\mathrm{m},0\mathrm{m},0\mathrm{m)}$ with the side length of $10$ m. For all the three configurations of numbers of semi-passive elements, the proposed multi-user location sensing algorithm achieves a high positioning accuracy of at least 3 cm with the probability of 90\%.
Moreover, by increasing the number of semi-passive elements to $400$, a millimeter-level localization accuracy is achieved with the probability of 90\%, due to a high spatial resolution.
\subsection{Performance of the Proposed Sensing-Based Beamforming Algorithms}
In this subsection, numerical results are presented to illustrate the effectiveness of the proposed beamforming algorithms in the ISAC and PC periods, respectively. For comparison, the alternating optimization (AO) algorithm \cite{8982186} with continuous phase shifts and perfect CSI  and the random phase shift algorithm are presented as two benchmarks. The sum rates in the ISAC and PC periods are respectively defined as
\begin{align}
&\bar{R}_{\mathrm{ISAC}}=\mathbb{E} \left\{ R\left( t \right) \right\} ,t\in \left\{ 1,2,\cdots ,T_1 \right\} ,\\
&\bar{R}_{\mathrm{PC}}=\mathbb{E} \left\{ R\left( t \right) \right\} ,t\in \left\{ T_1+1,T_1+2,\cdots ,T_1+T_2 \right\}.
\end{align}

\vspace{-0.6cm}
\begin{figure}[htbp]
  \centering
  \setlength{\abovecaptionskip}{-0.07cm}
  \setlength{\belowcaptionskip}{0.07cm}
  \subfigure[ISAC period.]
  {
  \label{trans_ISAC_M1}
  \includegraphics[width=3.2in]{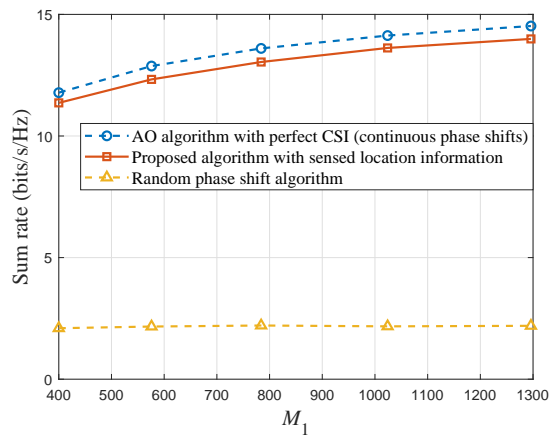}
  }
  \subfigure[PC period.]
  {
  \label{trans_PC_M1}
  \includegraphics[width=3.2in]{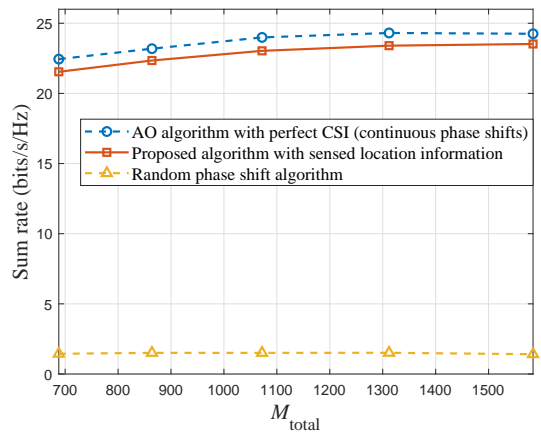}
  }
  \caption{Sum rate versus the number of IRS elements operating in the reflecting mode.}
\end{figure}

Fig.~\ref{trans_ISAC_M1} and Fig.~\ref{trans_PC_M1} compare the sum rates of different beamforming algorithms versus the number of IRS elements operating in the reflecting mode. It is obvious that the proposed algorithm with discrete phase shifts and sensed location information is superior to the random phase shift algorithm, and even achieves comparable performance to the AO algorithm with continuous phase shifts and perfect CSI. Moreover, with the increase of the number of IRS elements, all beamforming algorithms except for the random phase shift algorithm achieve significant performance gains, especially in the case of a small number of IRS elements. 
Even with the same number of IRS elements operating in the reflecting mode, the sum rate achieved in the PC period is much larger than that achieved in the ISAC period. This is because there are more distributed sub-IRSs assisting uplink data transmission in the PC period, thereby providing a larger spatial multiplexing gain.
\vspace{-0.4cm}
\begin{figure}[htbp]
  \centering
  \subfigure[ISAC period.]
  {
  \label{trans_ISAC_K}
  \includegraphics[width=3.2in]{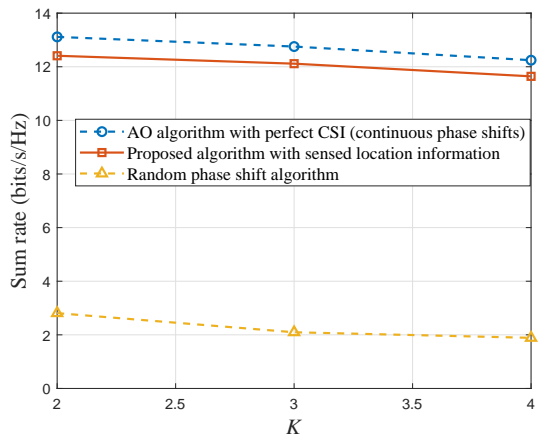}
  }
  \subfigure[PC period.]
  {
  \label{trans_PC_K}
  \includegraphics[width=3.2in]{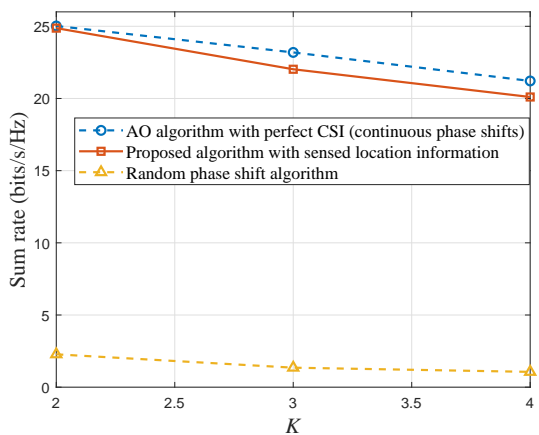}
  }
  \caption{Sum rate versus the number of users $K$.}
\end{figure}

Fig.~\ref{trans_ISAC_K} and Fig.~\ref{trans_PC_K} compare the sum rates of different beamforming algorithms versus the user number $K$, where we set the total transmit power of $K$ users to be $20$ dBm (i.e., $K\rho =20$ dBm) and $M_{\mathrm{semi}}=16$. In both ISAC and PC periods, the performance of all three beamforming algorithms degrades with the increase of the user number. For the AO algorithm and the random phase shift algorithm, this performance degradation is due to more inter-user interference, while for the proposed beamforming algorithm, this performance degradation is due to both more inter-user interference and less accurate sensed users' locations.
\vspace{-0.2cm}
\subsection{Performance of the IRS-Enabled Multi-User ISAC System}
\begin{figure}[htbp]
  \centering
  \setlength{\abovecaptionskip}{-0.1cm}
  \includegraphics[width=3.2in]{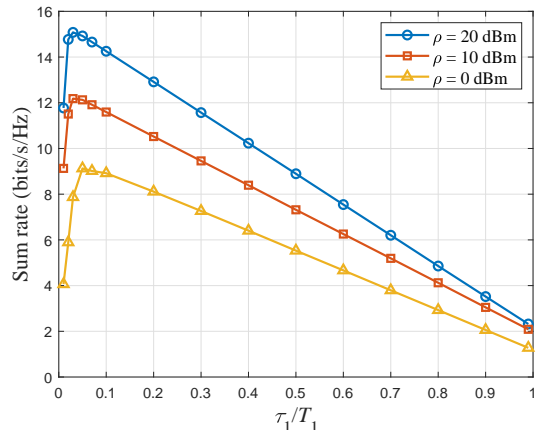}
  \caption{Sum rate of the ISAC period versus $\tau _1/T_1$.}
  \label{trans_ISAC_ratio}
  \vspace{-0.25cm}
\end{figure}

Fig.~\ref{trans_ISAC_ratio} illustrates the sum rate in the ISAC period versus $\tau _1/T_1$ with different values of $\rho$, where we set $T_1:T=\frac{1}{10}$ and $T=2000$. The optimal ratio of time allocation indicates that, it is desirable to allocate a small portion of time slots to the first time block and more time slots to the second time block. This is because the CSI is unavailable in the first time block, while location information is available in the second time block, which can be used for more effective beamforming design. Moreover, the optimal ratio $\tau_1/T_1$ decreases with the transmit power. With a larger transmit power, the positioning accuracy is sufficiently high and thus it is unnecessary to allocate too much time to the first time block.
However, with too little time allocated to the first time block, the communication performance would degrade significantly, due to the fact that low-accuracy localization in the first time block would affect the effectiveness of beamforming design in the second time block.
\vspace{-0.2cm}
\begin{figure}[!h]
  \centering
  \setlength{\abovecaptionskip}{-0.1cm}
  \includegraphics[width=3.2in]{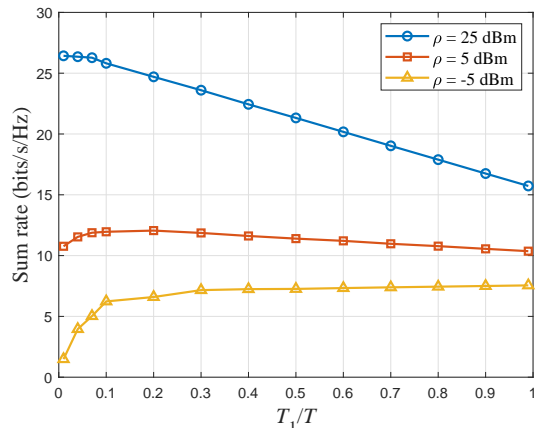}
  \caption{Sum rate of the ISAC period versus $T_1/T$.}
  \label{trans_PC_ratio}
\end{figure}

Finally, Fig.~\ref{trans_PC_ratio} presents the sum rate in the whole transmission period versus the ratio $T_1/T$ with different $\rho$, where $\tau _1/T_1=\frac{1}{10}$, $T=1000$, and the sum rate is defined as
\begin{align}
\bar{R}=\mathbb{E} \left\{ R\left( t \right) \right\} ,t\in \left\{ 1,2,\cdots ,T_1+T_2 \right\}.
\end{align}
The optimal ratio decreases as the transmit power becomes larger. For instance, with a lower power of $-5$ dBm, the optimal $T_1/T$ is about $1$, which indicates that simultaneous location sensing and communication should be conducted during the whole transmission period. As the transmit power increases to 
$5$ dBm, the optimal $T_1/T$ drops to $0.2$. With the highest transmit power of $25$ dBm, the optimal ratio approaches $0$. This is because with high transmit power, an enough high sensing accuracy can be achieved in a short time.

\section{conclusion}\label{section6}
In this paper, we proposed an IRS-enabled multi-user ISAC framework, and designed its working process, including the transmission protocol, multi-user location sensing as well as joint active and passive beamforming. Specifically, we first designed an ISAC transmission protocol, where the considered transmission period consists of the ISAC period for multi-user simultaneous communication and localization and the PC period for uplink data transmission. Then, we proposed a multi-user location sensing algorithm, by using  uplink communication signals sent by multiple users to the BS. Based on the sensed users' locations,  two novel beamforming algorithms were proposed to maximize the sum rate for the ISAC period and the PC period, respectively. Numerical results 
demonstrated that the proposed multi-user location sensing algorithm can provide millimeter-level accuracy with a high probability of $90\%$. By increasing the number of semi-passive elements or sensing time, the positioning accuracy can be further improved. Moreover, despite imperfect CSI and discrete phase shifts, the proposed two sensing-based beamforming algorithms achieve comparable performance to the benchmark  with perfect CSI and continuous phase shifts, demonstrating the effectiveness of beamforming designs using sensed location information.
In addition, we have investigated the overall communication performance of the IRS-enabled ISAC system, and
 numerical results showed that the optimal time  ratio of the ISAC period to the PC period decreases with the increase of transmit power, indicating that less time should be allocated to the ISAC period, when transmit power is high.
\bibliographystyle{IEEEtran}
\bibliography{references}{}

\begin{thebibliography}{10}
\providecommand{\url}[1]{#1}
\csname url@samestyle\endcsname
\providecommand{\newblock}{\relax}
\providecommand{\bibinfo}[2]{#2}
\providecommand{\BIBentrySTDinterwordspacing}{\spaceskip=0pt\relax}
\providecommand{\BIBentryALTinterwordstretchfactor}{4}
\providecommand{\BIBentryALTinterwordspacing}{\spaceskip=\fontdimen2\font plus
\BIBentryALTinterwordstretchfactor\fontdimen3\font minus
  \fontdimen4\font\relax}
\providecommand{\BIBforeignlanguage}[2]{{%
\expandafter\ifx\csname l@#1\endcsname\relax
\typeout{** WARNING: IEEEtran.bst: No hyphenation pattern has been}%
\typeout{** loaded for the language `#1'. Using the pattern for}%
\typeout{** the default language instead.}%
\else
\language=\csname l@#1\endcsname
\fi
#2}}
\providecommand{\BIBdecl}{\relax}
\BIBdecl

\bibitem{9424177}
Y.~Liu, X.~Liu, X.~Mu, T.~Hou, J.~Xu, M.~Di~Renzo, and N.~Al-Dhahir,
  ``Reconfigurable intelligent surfaces: Principles and opportunities,''
  \emph{IEEE Commun. Surveys Tuts.}, vol.~23, no.~3, pp. 1546--1577, May 2021.

\bibitem{wu2019intelligent}
Q.~Wu and R.~Zhang, ``Intelligent reflecting surface enhanced wireless network
  via joint active and passive beamforming,'' \emph{IEEE Trans. Wireless
  Commun.}, vol.~18, no.~11, pp. 5394--5409, Aug. 2019.

\bibitem{gong2020toward}
S.~Gong, X.~Lu, D.~T. Hoang, D.~Niyato, L.~Shu, D.~I. Kim, and Y.-C. Liang,
  ``Toward smart wireless communications via intelligent reflecting surfaces: A
  contemporary survey,'' \emph{IEEE Commun. Surveys Tuts.}, vol.~22, no.~4, pp.
  2283--2314, Jun. 2020.

\bibitem{wu2019towards}
Q.~Wu and R.~Zhang, ``Towards smart and reconfigurable environment: Intelligent
  reflecting surface aided wireless network,'' \emph{IEEE Commun. Mag.},
  vol.~58, no.~1, pp. 106--112, Nov. 2019.

\bibitem{pan2021reconfigurable}
C.~Pan, H.~Ren, K.~Wang, J.~F. Kolb, M.~Elkashlan, M.~Chen, M.~Di~Renzo,
  Y.~Hao, J.~Wang, A.~L. Swindlehurst, X.~You, and L.~Hanzo, ``Reconfigurable
  intelligent surfaces for 6{G} systems: Principles, applications, and research
  directions,'' \emph{IEEE Commun. Mag.}, vol.~59, no.~6, pp. 14--20, Jul.
  2021.

\bibitem{you2020channel}
C.~You, B.~Zheng, and R.~Zhang, ``Channel estimation and passive beamforming
  for intelligent reflecting surface: Discrete phase shift and progressive
  refinement,'' \emph{IEEE J. Select. Areas Commun.}, vol.~38, no.~11, pp.
  2604--2620, Jul. 2020.

\bibitem{yu2019miso}
X.~Yu, D.~Xu, and R.~Schober, ``{MISO} wireless communication systems via
  intelligent reflecting surfaces,'' in \emph{Proc. 2019 IEEE/CIC Int. Conf.
  Commun. China (ICCC), Changchun, China}, 2019, pp. 735--740.

\bibitem{zhou2020spectral}
S.~Zhou, W.~Xu, K.~Wang, M.~Di~Renzo, and M.-S. Alouini, ``Spectral and energy
  efficiency of {IRS}-assisted {MISO} communication with hardware
  impairments,'' \emph{IEEE Wireless Commun. Lett.}, vol.~9, no.~9, pp.
  1366--1369, Apr. 2020.

\bibitem{you2020energy}
L.~You, J.~Xiong, D.~W.~K. Ng, C.~Yuen, W.~Wang, and X.~Gao, ``Energy
  efficiency and spectral efficiency tradeoff in {RIS}-aided multiuser {MIMO}
  uplink transmission,'' \emph{IEEE Trans. Signal Process.}, vol.~69, pp.
  1407--1421, Dec. 2020.

\bibitem{pan2020intelligent}
C.~Pan, H.~Ren, K.~Wang, M.~Elkashlan, A.~Nallanathan, J.~Wang, and L.~Hanzo,
  ``Intelligent reflecting surface aided {MIMO} broadcasting for simultaneous
  wireless information and power transfer,'' \emph{IEEE J. Select. Areas
  Commun.}, vol.~38, no.~8, pp. 1719--1734, Jun. 2020.

\bibitem{zhou2020intelligent}
G.~Zhou, C.~Pan, H.~Ren, K.~Wang, and A.~Nallanathan, ``Intelligent reflecting
  surface aided multigroup multicast {MISO} communication systems,'' \emph{IEEE
  Trans. Signal Process.}, vol.~68, pp. 3236--3251, Apr. 2020.

\bibitem{guo2020weighted}
H.~Guo, Y.-C. Liang, J.~Chen, and E.~G. Larsson, ``Weighted sum-rate
  maximization for reconfigurable intelligent surface aided wireless
  networks,'' \emph{IEEE Trans. Wireless Commun.}, vol.~19, no.~5, pp.
  3064--3076, Feb. 2020.

\bibitem{wang2019sum}
S.~Wang, Q.~Li, S.~X. Wu, and J.~Lin, ``Sum rate maximization for multiuser
  {MISO} downlink with intelligent reflecting surface,''
  \emph{arXiv:1912.09315}.

\bibitem{8849960}
W.~Chen, X.~Ma, Z.~Li, and N.~Kuang, ``Sum-rate maximization for intelligent
  reflecting surface based terahertz communication systems,'' in \emph{Proc.
  2019 IEEE/CIC Int. Conf. Commun. Workshops China (ICCC Workshops), Changchun,
  China}, 2019, pp. 153--157.

\bibitem{huang2019reconfigurable}
C.~Huang, A.~Zappone, G.~C. Alexandropoulos, M.~Debbah, and C.~Yuen,
  ``Reconfigurable intelligent surfaces for energy efficiency in wireless
  communication,'' \emph{IEEE Trans. Wireless Commun.}, vol.~18, no.~8, pp.
  4157--4170, Jun. 2019.

\bibitem{zeng2021energy}
M.~Zeng, E.~Bedeer, O.~A. Dobre, P.~Fortier, Q.-V. Pham, and W.~Hao,
  ``Energy-efficient resource allocation for {IRS}-assisted multi-antenna
  uplink systems,'' \emph{IEEE Wireless Commun. Lett.}, Mar. 2021.

\bibitem{forouzanmehr2021energy}
M.~Forouzanmehr, S.~Akhlaghi, A.~Khalili, and Q.~Wu, ``Energy efficiency
  maximization for {IRS}-assisted uplink systems: Joint resource allocation and
  beamforming design,'' \emph{IEEE Commun. Lett.}, vol.~25, no.~12, pp.
  3932--3936, Sep. 2021.

\bibitem{fang2020energy}
F.~Fang, Y.~Xu, Q.-V. Pham, and Z.~Ding, ``Energy-efficient design of
  {IRS-NOMA} networks,'' \emph{IEEE Trans. Veh. Technol.}, vol.~69, no.~11, pp.
  14\,088--14\,092, Sep. 2020.

\bibitem{hu2018beyond}
S.~Hu, F.~Rusek, and O.~Edfors, ``Beyond massive {MIMO}: The potential of
  positioning with large intelligent surfaces,'' \emph{IEEE Trans. Signal
  Process.}, vol.~66, no.~7, pp. 1761--1774, Jan. 2018.

\bibitem{he2020large}
J.~He, H.~Wymeersch, L.~Kong, O.~Silv{\'e}n, and M.~Juntti, ``Large intelligent
  surface for positioning in millimeter wave {MIMO} systems,'' in \emph{Proc.
  2020 IEEE 91st Veh. Technol. Conf. (VTC2020-Spring), Antwerp, Belgium}, 2020,
  pp. 1--5.

\bibitem{he2020adaptive}
J.~He, H.~Wymeersch, T.~Sanguanpuak, O.~Silv{\'e}n, and M.~Juntti, ``Adaptive
  beamforming design for mmwave {RIS}-aided joint localization and
  communication,'' in \emph{Proc. 2020 IEEE Wireless Commun. Netw. Conf.
  Workshops (WCNCW), Seoul, Korea}, 2020, pp. 1--6.

\bibitem{elzanaty2021reconfigurable}
A.~Elzanaty, A.~Guerra, F.~Guidi, and M.-S. Alouini, ``Reconfigurable
  intelligent surfaces for localization: Position and orientation error
  bounds,'' \emph{IEEE Trans. Signal Process.}, vol.~69, pp. 5386--5402, Aug.
  2021.

\bibitem{wang2021joint}
W.~Wang and W.~Zhang, ``Joint beam training and positioning for intelligent
  reflecting surfaces assisted millimeter wave communications,'' \emph{IEEE
  Trans. Wireless Commun.}, vol.~20, no.~10, pp. 6282--6297, Apr. 2021.

\bibitem{zhang2020towards}
H.~Zhang, H.~Zhang, B.~Di, K.~Bian, Z.~Han, and L.~Song, ``Towards ubiquitous
  positioning by leveraging reconfigurable intelligent surface,'' \emph{IEEE
  Commun. Lett.}, vol.~25, no.~1, pp. 284--288, Sep. 2020.

\bibitem{zhang2021metalocalization}
H.~\vspace{0mm}Zhang, H.~Zhang, B.~Di, K.~Bian, Z.~Han, and L.~Song,
  ``Metalocalization: Reconfigurable intelligent surface aided multi-user
  wireless indoor localization,'' \emph{IEEE Trans. Wireless Commun.}, vol.~20,
  no.~12, pp. 7743--7757, Jun. 2021.

\bibitem{wang2021joint2}
R.~Wang, Z.~Xing, and E.~Liu, ``Joint location and communication study for
  intelligent reflecting surface aided wireless communication system,''
  \emph{arXiv:2103.01063}.

\bibitem{9235486}
J.~Yuan, Y.-C. Liang, J.~Joung, G.~Feng, and E.~G. Larsson, ``Intelligent
  reflecting surface-assisted cognitive radio system,'' \emph{IEEE Trans.
  Commun.}, vol.~69, no.~1, pp. 675--687, Oct. 2021.

\bibitem{jung2011wi}
S.~Jung, C.~oh~Lee, and D.~Han, ``{Wi-Fi} fingerprint-based approaches
  following log-distance path loss model for indoor positioning,'' in
  \emph{Proc. 2011 IEEE MTT-S Int. Microw. Workshop Ser. Intell. Radio Future
  Pers. Terminals, Daejeon}, 2011, pp. 1--2.

\bibitem{5535154}
X.~Chen and Z.~Zhang, ``Exploiting channel angular domain information for
  precoder design in distributed antenna systems,'' \emph{IEEE Trans. Signal
  Process.}, vol.~58, no.~11, pp. 5791--5801, Aug. 2010.

\bibitem{8982186}
H.~Guo, Y.-C. Liang, J.~Chen, and E.~G. Larsson, ``Weighted sum-rate
  maximization for reconfigurable intelligent surface aided wireless
  networks,'' \emph{IEEE Trans. Wireless Commun.}, vol.~19, no.~5, pp.
  3064--3076, Feb. 2020.

\end{thebibliography}

\end{document}